\documentclass[twocolumn]{autart}    

\usepackage{graphicx}          

\usepackage{amsmath}

\usepackage{amsfonts}
\usepackage{color}
\usepackage{epstopdf} 
\usepackage{tikz}
\usepackage{mathtools}
\usepackage{soul}

\newtheorem{proposition}{Proposition}

\newtheorem{remark}{Remark}
\newtheorem{lemma}{Lemma}
\newtheorem{assumption}{Assumption}
\newtheorem{Algorithm}{Algorithm}

\newcommand{\field}[1]{\mathbb{#1}}
\newcommand{\R}{\field{R}}

\newcommand{\beq}{\begin{equation}}
\newcommand{\eeq}{\end{equation}}
\newcommand{\beqs}{\begin{equation*}}
\newcommand{\eeqs}{\end{equation*}}
\newcommand{\beqr}{\begin{eqnarray}}
\newcommand{\eeqr}{\end{eqnarray}}
\newcommand{\beqrs}{\begin{eqnarray*}}
	\newcommand{\eeqrs}{\end{eqnarray*}}
\newcommand*{\blue}{\textcolor{black}}

\usepackage{algorithm}

\newcommand{\zn}{\hat\zeta^n_N}
\newcommand{\zo}{\zeta^0}
\newcommand{\etaNj}{\hat\eta^n_{j_N}}

\usepackage{cite}
\usepackage{enumitem}

\begin{document}

\begin{frontmatter}

\title{A scalable multi-step least squares method for network identification with unknown disturbance topology\thanksref{footnoteinfo}} 
\thanks[footnoteinfo]{Paper submitted to Automatica, 14 June 2021. \blue{Revised 23 November 2021.} Final version 28 January 2022. Corrected 4 April 2022. This project has received funding from the European Research Council (ERC), Advanced Research Grant SYSDYNET, under the European Union’s Horizon 2020 research and innovation programme (Grant Agreement No. 694504).}

\author[Paestum]{Stefanie J.M. Fonken}\ead{s.j.m.fonken@tue.nl},               
\author[Paestum]{Karthik R. Ramaswamy}\ead{k.r.ramaswamy@tue.nl},    
\author[Paestum]{Paul M.J. Van den Hof}\ead{p.m.j.vandenhof@tue.nl}  

\address[Paestum]{Department of Electrical Engineering, Eindhoven University of Technology, Eindhoven, The Netherlands}


\begin{abstract}
Identification methods for dynamic networks typically require prior knowledge of the network and disturbance topology, and often rely on solving poorly scalable non-convex optimization problems. While methods for estimating network topology are available in the literature, less attention has been paid to estimating the disturbance topology, i.e., the (spatial) noise correlation structure and the noise rank \blue{in a filtered white noise representation of the disturbance signal}.
In this work we present an identification method for dynamic networks, in which an estimation of the disturbance topology precedes the identification of the full dynamic network with known network topology.
To this end we extend the multi-step Sequential Linear Regression and Weighted Null Space Fitting methods to deal with reduced rank noise,
and use these methods to estimate the disturbance topology and the network dynamics \blue{in the full measurement situation}. As a result, we provide a multi-step least squares algorithm with parallel computation capabilities and that rely only on explicit analytical solutions, thereby avoiding the usual non-convex optimizations involved. Consequently we consistently estimate dynamic networks of Box Jenkins model structure, while keeping the computational burden low.
We provide a consistency proof that includes path-based data informativity conditions for allocation of excitation signals in the experimental design. Numerical simulations performed on a dynamic network with reduced rank noise clearly illustrate the potential of this method.
\end{abstract}
\end{frontmatter}
%
\section{Introduction}
Dynamic networks represent large-scale interconnected systems, and data-driven modeling of dynamic networks has received considerable attention in recent years.
These networks can be considered as a set of measurable (node) signals interconnected through linear dynamic systems (the modules), driven by measured external excitation signals and/or unmeasured disturbance signals.
Modeling of these networks plays an important role in biological systems \cite{Bio_net,brain_netw}, economic systems\cite{materassi2010topological}, power  networks \cite{pagani_powergrid}, and many other fields in science and engineering. 
The challenges addressed in identification of dynamic networks can roughly be divided into three categories.
The first is identifying the interconnection structure of the nodes in a dynamic network referred to as network topology detection \cite{materassi2010topological,chiuso2012bayesian}. The second is the identification of a specific module in a network, referred to as local module identification. For this problem
 closed-loop identification methods have been generalized to the dynamic network situation in \cite{VandenHof2013}, formulating the local module identification problem as a multi-input-single-output (MISO) problem. This has been further extended and generalized in e.g., \cite{Dankers:15EIV,morsm2016,Dankers&etal_TAC:16,Gevers&etal_sysid:18,Materassi&Salapaka:20,Ramaswamy&etal_Autom:21,Ramaswamy&etal_TAC:21,hof_karthik_sur:21a}. The third challenge is identification of the full network dynamics \cite{pem_joint_direct,Harm_SLS,Dankers_SLR,WNSF_stef}, where the problem is formulated as the identification of a (structured) multi-input-multi-output (MIMO) model.

In this paper we will further explore the development of full network identification methods.
While dynamic networks increase in complexity and size, and measurement data is becoming  increasingly accessible, there is a strong demand for accurate and scalable data driven modeling methods.
The joint direct method \cite{weerts_rank_reduced,pem_joint_direct} predicts all node signals in the network jointly and achieves consistency and minimum variance properties in the situation that the network and disturbance topology are given a priori and the noise can be of reduced rank. However it strongly relies on solving (constrained) non-convex optimization problems, which seriously limits its scalability to
larger networks.
There are multi-step convex identification methods available for full network identification, such as the Sequential Linear Regression (SLR) \cite{Dankers_SLR}, Sequential Least Squares (SLS) \cite{Harm_SLS} and extensions of Weighted Null Space Fitting (WNSF) \cite{Galrinho_Para_WNSF_2019} such as \cite{WNSF_stef}.
Moreover, methods such as the SLR and SLS allow for splitting the MIMO optimization into \blue{multiple linear regressions}, which contributes to a lower computational burden.
The available convex methods are scalable to larger networks, but are limited to particular model structures of the network, and additionally, they do not allow for handling reduced rank noise. Particularly in large-scale network identification,
stepping away from the typical assumption that all disturbance signals have their own  independent noise source, is an appealing situation that should be supported by an effective estimation algorithm. Handling this situation of reduced-rank noise can substantially reduce the variance of estimated models. However it also introduces the problems of estimating the noise rank and noise correlation structure from data.

All available convex and non-convex methods for network identification require prior knowledge on the topology (i.e. rank and spatial
correlation structure of the disturbance model). While in dynamic factor analysis \cite{Deistler2015} attention has been paid to the estimation of noise rank, in prediction error identification this does not appear to be included yet in the identification algorithms.
For situations where the disturbance topology information is not readily available, it is attractive to develop methods that include estimating this information from data.

The topology estimation literature shows a variety of available methods to estimate the topology, such as
Wiener filter based methods \cite{materassi2010topological,Materassi_Relations,materassi2012problem},
Bayesian model selection techniques \cite{WASSERMAN200092,chiuso2012bayesian,shi2019bayesian}, or methods that infer the topology from parametric estimates \cite{bolstad2011causal,yuan2011robust,DANKERS2012structure}.
\blue{While the main focus of topology detection literature has been on estimating network topology in the situation of a diagonal disturbance spectrum $\Phi_v(\omega)$, extensions towards nondiagonal spectra have been presented in \cite{Dimovska2017,veedu2020topology,BOMBOIS_hal_2021}. In \cite{veedu2020topology} network topology and the non-zero pattern in the disturbance spectrum are estimated jointly.}
In this paper we assume that we do not know the disturbance topology a priori, but we assume that the network topology is known \blue{e.g., from its underlying physics, which is commonly the case for engineered systems. In the situation that the network topology is not known beforehand, it is possible to use any of the above cited methods to estimate it.}
We allow the process noise to be spatially correlated, i.e. the disturbance spectrum $\Phi_v(\omega)$ is not necessarily diagonal. Additionally the noise is allowed to be of reduced rank, i.e. $\Phi_v(\omega)$ can be singular.

The objective is to develop a multi-step convex algorithm that estimates the disturbance topology and the dynamic modules in the network for general model structures including the Box Jenkins (BJ) structure, while adhering to computational algorithms that are scalable, while achieving favorable properties in terms of low experiment cost, consistency and reduced variance of the network estimates.

To this end we develop a multi-step algorithm to identify the network dynamics. In the first step the noise rank and the
\blue{nonzero pattern in the corresponding disturbance model (noise shaping filter) are estimated.}
This is done through a (nonparametric) high-order ARX model, inspired by the SLR method \cite{Dankers_SLR}. Next, this information is used to develop a multi-step convex algorithm that can accurately identify the dynamics of the network in the situation of reduced rank noise and for a very general Box Jenkins model structure, thereby combining the recently introduced multi-step convex identification methods SLR \cite{Dankers_SLR} and WNSF \cite{Galrinho_Para_WNSF_2019,WNSF_stef} and extending them to the described situation.

The paper proceeds with a definition of the considered dynamic network setup in Section \ref{sec:dyn_net}. In Section \ref{sec:noise_top} we present a new method for estimating the disturbance topology from data, followed in Section \ref{sec:Id} by a multi-step identification algorithm that exploits the prior estimated disturbance topology.
Section \ref{sec:Analyses} presents the consistency analysis of the method, including graph-based conditions for data informativity. Results of  numerical simulations are provided in Section \ref{sec:Num_ex}, followed by  conclusions in Section \ref{sec:conclusion}. The consistency proofs are collected in the Appendix.
\section{Dynamic networks}\label{sec:dyn_net}
Following the setting of \cite{VandenHof2013} a dynamic network is defined by $L$ nodes or internal variables $w_j(t), \, j=1, \dots, L$, that are scalar-valued measured signals. The underlying network is linear time invariant (LTI), and the nodes of the network can be expressed as
\begin{equation}
    w_j (t) = \sum_{\substack{l\in \mathcal{N}_j}} G^0_{jl}(q) w_l(t) + \sum_{\substack{k\in \mathcal{R}_j}} R^0_{jk}(q)r_k(t) +v_j(t),
    \label{eq:dyn_network1}
\end{equation}
where
\begin{itemize}
    \item $q^{-1}$ the delay operator, i.e. $q^{-1} w_j(t)=w_j(t-1)$,
    \item $\mathcal{N}_j$ defines the set of indices of measured node signals $w_l$, $l\neq j$, for which $G^0_{jl}(q) \neq 0$, where $G^0_{jl}(q)$ is a strictly proper rational transfer function,
    \item $\mathcal{R}_j$ defines the set of indices of measured external excitation signals $r_k$, for which $R^0_{jk}(q) \neq 0$, where $R^0_{jk}(q)$ is a known proper rational transfer function, 
    \item $v_j(t)$ is unmeasured process noise, where the disturbance vector $v=[v_1 \cdots v_L]^\top$ is modeled as a \blue{wide sense} stationary stochastic process represented by $v(t) = H^0(q) e(t)$. The $e = [e_1 \cdots e_p]^\top$ is a white noise process of \blue{dimension} $p\leq L$ with covariance matrix $\Lambda^0 > 0$.  $H^0(q)$ is a rational transfer function matrix.
\end{itemize}
The full network expression, with omitted $q$ and $t$, is
\begin{equation}
    \begin{bmatrix}
    w_1 \\ w_2 \\ \vdots \\w_L
    \end{bmatrix} \!\!=\!\!
    \begin{bmatrix}
    0 &G_{12}^0 & \cdots & G_{1L}^0 \\
    G_{21}^0 & 0 & \ddots & G_{2L}^0\\
    \vdots & \ddots &\ddots &\vdots \\
    G_{L1}^0 & G_{L2}^0 & \cdots & 0
    \end{bmatrix}
    \!\!
    \begin{bmatrix}
    w_1 \\ w_2 \\ \vdots \\w_L
    \end{bmatrix} \!\! + \!
    R^0 \!\!
    \begin{bmatrix}
    r_1 \\ r_2 \\ \vdots \\r_K
    \end{bmatrix} \!\! + \!
    H^0 \!\!
    \begin{bmatrix}
    e_1 \\ e_2 \\ \vdots \\e_p
    \end{bmatrix}
    \label{eq:dyn_netw_MIMO}
\end{equation}
with the matrix notation given by
\begin{subequations}
\begin{equation}
    w  =G^0 w+ R^0 r+H^0 e , \label{eq:dyn_model_a}
\end{equation}
    \begin{equation}
    w  = (I-G^0)^{-1} (R^0r+H^0e),
    \label{eq:dyn_model_b}
    \end{equation}
\end{subequations}
where we assume that the inverse $(I-G^0)^{-1}$ exists and the network is well-posed, as used in \cite{VandenHof2013}.\\
In the situation $p < L$, i.e. when the noise is of reduced rank or singular, the disturbance model $H^0$ is a non-square matrix, i.e.
\begin{itemize}
     \item $H^0 \in \mathbb{R}^{L\times p}(z)$ is stable and has a stable left inverse $H^\dagger$ 
     that satisfies $H^\dagger H=I \in \mathbb{R}^{p\times p}$; 
\end{itemize}
%
For a unique representation of reduced rank spectra that can be used to construct a predictor we can adopt a result from \cite{pem_joint_direct} where the disturbance term is equivalently written as $\breve H^0 \breve e$ with $\breve H^0$ square.
\begin{lemma}[\cite{pem_joint_direct}]
\label{lem1}
Consider an $L$-dimensional disturbance process $v$ with rank $p$. Then the disturbance signals $v$ can be reordered in such a way that the following unique representations result:
\begin{equation}
\begin{aligned}
   &\begin{bmatrix} v_a \\ v_b  \end{bmatrix} = H^0 e=
         \breve{H}^0 \breve{e} \quad \text{ with }\\
         & H^0= \begin{bmatrix}
    H_a^0 \\ H_b^0
    \end{bmatrix},\
    \breve{H}^0=
    \begin{bmatrix}
    H_a^0  & 0\\ H_b^0-\Gamma^0 & I
    \end{bmatrix}, \
    \breve{e}=\begin{bmatrix} \breve e_a \\  \breve e_b  \end{bmatrix}
    \!=\!\begin{bmatrix} e \\ \Gamma^0 e  \end{bmatrix} \\
    & \mbox{and } \Gamma^0 = \text{lim}_{z\to \infty}H_b^0 (z)
    \label{eq:def_H_Hbreve}
    \end{aligned}
\end{equation}
such that
\begin{itemize}
    \item $H_a^0 \in \mathbb{R}^{p\times p}(z)$ is a monic full rank rational transfer function matrix;
    \item $H_b^0\in \mathbb{R}^{(L-p)\times p}(z)$ is a stable proper rational transfer function matrix.
    \item The covariance matrix of $\breve e$ is given by,
    \begin{equation}
    \breve{\Lambda}^0 =
    \begin{bmatrix}
    I \\ \Gamma^0
    \end{bmatrix}
    \Lambda^0
        \begin{bmatrix}
    I \\ \Gamma^0
    \end{bmatrix}^\top =
    \begin{bmatrix}
    \Lambda^0 & \Lambda^0 \Gamma^{0^\top} \\
    \Gamma^0 \Lambda^0 & \Gamma^0 \Lambda^0 \Gamma^{0^\top}
    \end{bmatrix},
    \label{eq:Lambda_breve}
\end{equation}
where $\Lambda^0 \in \mathbb{R}^{p\times p}$ has rank $p$. \hfill $\Box$
\item If additionally $H_a^0$ is minimum phase then $\breve H^0$ is monic, stable and minimum phase.\footnote{It has recently been pointed out in \cite{Cao&Picci&Lindquist_id:21} that this excludes the situation where the (deterministic) mapping from $v_a$ to $v_b$ is unstable.}
\end{itemize}
\end{lemma}
The result of the reordering of signals as indicated in the Lemma is that the first $p$ components of the reordered signal constitute a full rank $p$ process.

We assume that the data generating network satisfies the following properties.
\begin{assumption} \mbox{ }
\begin{enumerate}[label=\alph*.]
    \item The network is well-posed, i.e. all principle minors of $\big(I-G^0 (\infty)\big)$ are nonzero \cite{Araki}. 
    \item $(I-G^0)^{-1}$ is stable and causal.
    \item All elements in $G(q)$ are strictly proper.
    \item $H^0$ is stable and has a stable left inverse.
    \item $\breve H^0$ is square, monic and minimum phase.
    \item The topology of $G^0$ and $R^0$, and \blue{the non-zero elements of} $R^0$ are fixed and known.
    \item The matrix $R^0$ has a block diagonal structure:  $R^0 = diag(R^0_a, R^0_b)$ in the situation of ordered nodes as meant in \eqref{eq:def_H_Hbreve}.
    \item {Measurements} of {all} node signals $w$ and all present excitation signals $r$ are available.
    \item \blue{The standard regularity conditions on the data are satisfied that are required for consistency results of the prediction error identification method.\footnote{\blue{See \cite{ljung99} page 249. This includes the property that $e(t)$ has bounded moments of order higher than 4.}}}
\end{enumerate}
\end{assumption}

The two main steps of the identification method that will be developed in this paper are
\begin{itemize}
    \item Estimating the disturbance topology, i.e. the noise rank and the zero pattern in the disturbance model.
    \item Estimating the dynamical components in the network for a given network and disturbance topology, while using a parametric BJ model structure.
\end{itemize}
In the next section we first focus on the disturbance topology estimation method, followed by the developed identification method in the section thereafter.
\section{Disturbance topology estimation}\label{sec:noise_top}
%
Before we can use a unique disturbance model that is structured according to $\breve H^0$ in (\ref{eq:def_H_Hbreve}), we need to estimate the noise rank $p$ and we need to be able to reorder the node signals in such a way that a noise representation as in (\ref{eq:def_H_Hbreve}) can be used. This step is necessary as the unstructured disturbance model $H^0$ is non-unique in the situation $p < L$. Therefore the disturbance topology estimation is performed in two main steps:
\begin{itemize}
    \item Step 1: Estimating the noise rank, and reordering the signals to the situation of Lemma \ref{lem1}.
    \item Step 2: Estimating the structure of the disturbance model $\breve H^0$.
\end{itemize}
\subsection{Step 1: Estimating noise rank $p$ and reordering of nodes}
\label{sec:3.1 step1}
For estimating the noise rank $p$, we are going to estimate the covariance matrix $\breve \Lambda^0$ (\ref{eq:Lambda_breve}) of innovation signal $\breve e$, which through its rank $p$ can provide us access to the correct noise rank.\\
An estimate of the covariance matrix is obtained by estimating a high-order (nonparametric) ARX model on the basis of measured signals $w, r$, and by using the residual (predictor error) of this estimated model as an estimate of the white noise term $\breve e$.

A parametrized ARX model is chosen according to
\begin{eqnarray}
\breve  A(q,\zeta) & = & I +\breve A_1 q^{-1} + \cdots +\breve A_nq^{-n} \\
\breve  B(q,\zeta) & = &\breve B_0 + \breve B_1 q^{-1} + \cdots \breve B_{n\blue{-1}} q^{-(n\blue{-1})}
\end{eqnarray}
while all coefficients of \blue{$\breve A_k, \breve B_k$} are \blue{vectorized and} collected in the parameter vector $\zeta$.
The one-step-ahead predictor\blue{\cite{ljung99}, defined as}
\begin{equation}
    \hat{w}(t|t-1;\zeta) := {\mathbb{E}}\{ w(t)| w^{t-1}, \, r^{t}\},
    \label{eq:def_predictor_step1x}
\end{equation}
\blue{where $w^{t-1}$ and $r^{t}$ are defined according to $w^{t-1}:=\{w(0), w(1), \cdots, w(t-1)\}$ and $r^{t}:=\{r(0), r(1), \cdots, r(t)\}$,}
is given by
\begin{eqnarray}
    \hat{w}(t|t-1,\zeta) & = &\big(I- \breve{A}(\blue{q,}\zeta)\big) {w}(t) + \breve{B}(\blue{q,}\zeta) {r}(t)
    \label{eq:predictorARX1x} \\
    & = & \varphi(t) \zeta
\end{eqnarray}
with $\varphi(t)$ composed of the appropriate terms in $w$ and $r$.\\
Note that for an actual network with representation $G^0, \breve H^0, R^0$, the one-step predictor will be given by
\begin{eqnarray}
   \hat{w}(t|t-1) & =& \big(I- (\breve H^0(q))^{-1}(I-G^0(q))) w(t) +  \nonumber \\
   & & + (\breve H^0(q))^{-1}R^0(q) {r}(t).
   \label{eq:x1}
\end{eqnarray}
This implies that the polynomial predictor model (\ref{eq:predictorARX1x}) can only accurately approximate the rational filters that are present in (\ref{eq:x1}) if the ARX order $n$ is chosen very high. 
The ARX model is estimated according to $\hat \zeta^n_N = \arg\min_{\zeta} \frac{1}{N} \sum_{t=1}^N \varepsilon^T(t,\zeta)\varepsilon(t,\zeta)$, with $\varepsilon(t,\zeta) = w(t) - \hat w(t|t-1;\zeta)$, leading to the analytical solution
\begin{equation}
        \hat{\zeta}^n_{N}=
    \Bigg[ \frac{1}{N} \sum^N_{t=1} \varphi(t) \varphi^{\top}(t) \Bigg] ^{-1}
     \frac{1}{N} \sum^N_{t=1} \varphi(t)  {w}(t).
     \label{eq:zetax}
\end{equation}
%
\blue{Since the network identifiability conditions of \cite{WEERTS2018Identifiability_rr} are satisfied for the considered model set, the sample estimate}
\begin{equation}
        \hat{{\Lambda}} := \frac{1}{N} \sum^N_{t=1} {\varepsilon}(t,\hat{\zeta}^n_N) {\varepsilon}^\top(t,\hat{\zeta}^n_N),
        \label{eq:lambda_tildex}
\end{equation}
\blue{will then, under mild regularity conditions, be a consistent estimate of the noise covariance $\breve\Lambda^0$}.
\blue{The rank $p$ of the noise process can then be estimated through a rank test on $\hat\Lambda$, e.g., through a singular value decomposition. Alternatively, other matrix factorizations or information based criteria can be applied for estimating the rank, see e.g., \cite{CambaMendez&Kapetanios:09}.}
When $\hat\Lambda$ and \blue{the estimated} rank $\blue{\hat p} < L$ have been determined, the $L$ signals can be reordered through a 
permutation matrix $\Pi$ such  that the first $\blue{\hat p}$ components of the permuted noise vector have a rank $\blue{\hat p}$ covariance matrix, i.e.
$\begin{bmatrix} I_{\blue{\hat p}} & 0\end{bmatrix} \Pi^\top \hat{\Lambda} \Pi\begin{bmatrix} I_{\blue{\hat p}} & 0\end{bmatrix}^\top$ has rank $\blue{\hat p}$.
\begin{remark}
Since the polynomials $\breve A(\zeta)$ and $\breve B(\zeta)$ are fully parametrized with independent parameters on each polynomial entry, the MIMO least squares optimization that leads to the solution \eqref{eq:zetax} can also be decomposed in $L$ separate \blue{linear regressions that minimize the residual $\varepsilon_j(t,\zeta)$ separately for each $j$}, which is computationally attractive \blue{since the computations can be performed in parallel or sequentially}.
\end{remark}
\begin{remark}
The resulting estimation scheme will generally not provide us with consistent estimates of the ARX model. This is not only due to the fact that typically the order $n$ of the ARX model would need to go to infinity, \blue{but also to the fact that the solution for $\hat \zeta^n_N$ is non-unique in the situation $p<L$. However, this latter non-uniqueness does not affect the uniqueness and whiteness of the residual $\varepsilon(t,\zn)$ since, according to the projection theorem, every solution for $\hat \zeta^n_N$ determines the same predictor} \cite{Deistler2010}. 
\blue{The estimate $\hat \Lambda$ is therefore consistent, i.e. $ \hat \Lambda= cov(\breve e)$ w.p. 1 as $n, N \rightarrow \infty$}. 
\end{remark}
\blue{
\begin{remark}
Although a correct estimation of the noise rank $p$ cannot be guaranteed, consistency results for estimating $p$ would be possible when applying information-based criteria for rank estimation, e.g., based on the BIC criterion \cite{CambaMendez&Kapetanios:09}. In the next steps of our approach it will be assumed that a correct estimation of $p$ has been obtained.
\end{remark}
}
After reordering the node signals as described above, we can now adhere to a network representation with a unique disturbance model according to the structure in Lemma \ref{lem1}, where $\breve H^0$ can be parametrized by the transfer function \blue{matrices} $H_a$ and $H_b$.

\subsection{Step 2: Estimating the noise correlation structure}
\label{subsec:step2}
In the second step we are going to estimate which entries in our disturbance model are nonzero. To this end we extend the SLR method \cite{Dankers_SLR} to the situation of reduced rank noise and show how the noise correlation structure can be obtained.
\subsubsection{Step 2.1: Refining the nonparametric ARX model}
\label{subsec:consistency_zeta}
With the noise rank $p$ available and the nodes being ordered, we have gained additional information on $\breve{H}^0$ \eqref{eq:def_H_Hbreve}, namely the last $L-p$ columns are now known.
Now, we perform the same approach of identification using high order ARX modeling as in the previous step, but by utilizing the known entries in $\breve{H}^0$, leading to refined estimates of $\breve A(\zn)$ and $\breve B(\zn)$. In the analysis results of Section \ref{sec:anal1} it shown that the known entries in $\breve{H}^0$ can simply be mapped to known entries in the parametrized polynomial $\breve B(\zeta)$, and therefore can simply be taken into account in the least squares problem \eqref{eq:zetax}.
In Section \ref{sec:anal1} it is shown that this leads to consistent estimates $\zn$ for $n,\blue{N} \rightarrow \infty$.
\subsubsection{Step 2.2: Predictor model with reconstructed innovation input}
In this step we are going to use the estimated nonparametric ARX model to reconstruct the innovation signal. This allows us to use the reconstructed innovation signal as a measured input in the predictor model that will be used for estimating the structure of the disturbance model.

If there exists a parameter $\zo$ such that the ARX model $(\breve A(\zo), \breve B(\zo))$ captures the dynamics of the network, then it follows from \cite{pem_joint_direct} that
\begin{equation}
   \varepsilon(t,\zo) = \begin{bmatrix} I \\ \Gamma^0 \end{bmatrix} e(t).
\end{equation}
We can accordingly decompose $\varepsilon(t,\zeta)$ as
\begin{equation}
   \varepsilon(t,\zeta) = \begin{bmatrix} \varepsilon_a (t,\zeta) \\ \varepsilon_b (t,\zeta) \end{bmatrix}
      \label{eq:innovation_1}
\end{equation}
while the consistency property of $\zn$ implies that
\begin{equation}
\begin{aligned}
    \varepsilon_{a}(t,\hat{\zeta}^n_N) &\to e(t) \qquad \text{w.p. 1 as} \, N\to \infty \, \forall t ,\\
\varepsilon_{b}(t,\hat{\zeta}^n_N) &\to \Gamma^0 e(t) \quad \text{w.p. 1 as} \, N\to \infty \, \forall t .\\
\end{aligned}
\end{equation}
We will refer to $\varepsilon(t,\zn)$ as the ``reconstructed innovation''.\\
For a network with ordered nodes we evaluate a new one-step-ahead predictor
\begin{equation}
    \hat{w}(t|t-1) := {\mathbb{E}}\{ w(t)| w^{t-1}, \blue{r^{t}}, e^{t-1}\}
    \label{eq:def_pred_varepsilon_a}
\end{equation}
that includes the innovation signal \blue{$e^{t-1}:=\{e(0), e(1), \\ \cdots, e(t-1)\}$} in the expectation. Then it follows that
\begin{equation}
      \hat{w}(t|t-1) = G^0(q) w(t) + (\breve H^0(q)-I) \breve e(t) + R^0(q)r(t),
\end{equation}
where
\begin{equation}
\begin{aligned}
    (\breve{H}^0\!-\!I) \breve{e} &=
    \big( \begin{bmatrix}
        H_a^0 & 0 \\ H_b^0\! -\!\Gamma^0 & I
    \end{bmatrix}
    \! -\! I \big) \breve{e} = \begin{bmatrix}
        H_a^0\! -\! I  \\ H_b^0\! -\! \Gamma^0
    \end{bmatrix}
     e
     \blue{=\bar H^0 e}.
     \label{eq:Hbreve-I}
\end{aligned}
\end{equation}
This motivates the use of the following parametrized predictor model per node:
\begin{equation}
\begin{aligned}
    &\hat{w}_j(t|t-1,\eta_j)= \\
    \sum_{\substack{l\in \mathcal{N}_j}} 
   & G_{jl}(\eta_j) w_l  +
   \sum_{s\in \mathcal{V}_j}
      \bar{H}_{js} (\eta_j) \varepsilon_{a_s}(\zn)
   +
   \sum_{\substack{k\in \mathcal{R}_j}} R_{jk}r_k,
   \end{aligned}
   \label{eq:pred_step2_MISO}
\end{equation}
where the terms \blue{$G(\eta)$} and \blue{$\bar H(\eta)$} are parametrized \blue{versions of $G^0$ and $\bar H^0$ respectively, and $\varepsilon_{a}(\zn)$ is an estimate of the noise signal $e(t)$. $G_{jl}(\eta)= \sum_{k=1}^n g^{jl}_k q^{-k}$ and $\bar H_{js}(\eta)= \sum_{k=1}^n h^{js}_k q^{-k}$ are parametrized} as strictly proper polynomials of order $n$, the term $\sum_{\substack{k\in \mathcal{R}_j}} R_{jk}r_k(t)$ is known, the sets $\mathcal{N}_j$ and $\mathcal{R}_j$ are known from the topology of $G^0$ and $R^0$, and $\mathcal{V}_{j}$ defines the set of indices of noise signals for which noise dynamics is present in the disturbance model. This leads to an ARX model, like in Step 1, but now with the reconstructed innovation $\varepsilon_a(t,\zn)$ added as external predictor input signal, and the coefficients of the unknown polynomials collected in the parameter vector $\eta$. It is our next objective now to determine the sets $\mathcal{V}_j$ for $j = 1,\cdots, L$.
To this end we follow two approaches namely the structure selection approach and the Glasso approach, which will be presented next.

\subsubsection{Structure selection}
\label{sec:323}
For a particular choice of $\mathcal{V}_j$ we evaluate the residual $\varepsilon_j(t,\hat\eta^n_{N_j}) := w_j(t) - \hat w_j(t|t-1,\hat\eta^n_{N_j})$ where $\hat\eta^n_{N_j}$ is the estimated parameter that minimizes the quadratic criterion $\frac{1}{N}\sum_{t=1}^N \varepsilon_j^2(t,\eta_j)$, and that is obtained through an analytical solution, similar to \eqref{eq:zetax}. We test this residual with possible combinations in set $\mathcal{V}_j$ and employ model selection techniques such as AIC, BIC and {Cross-validation} (CV) on the obtained estimates $\hat{\eta}^n_{N_j}$ \cite{yuan2011robust}\blue{, of which the BIC provides a consistent estimate} \cite{schwarz1978estimating,Bayes_factors}.
Because we use ARX models to estimate $\eta$, model selection techniques such as AIC, BIC and CV are convex.
Additionally, since we derive the disturbance topology per node, we have to test at most $2^L$ possible sets $\mathcal{V}_j$ for $L$ nodes. This results in a lower computational burden compared to when we detect the topology in a MIMO setting, where we would have to test at most {$2^{L^2-L}$} possible sets $\mathcal{V}_j$ simultaneously for all $j$ \cite{yuan2011robust}.
However, for large networks these model selection techniques can still become computationally heavy.

\subsubsection{Sparse estimation with Glasso}
\label{sec:324}
For each node $j$, a Glasso (Group Lasso) estimate is computed by minimizing the following cost function over $\eta_j$ for a fully parametrized disturbance model with $p$ white noise inputs:
\begin{equation}
\min_{\eta_{j}} \left\{
     \frac{1}{2} \sum_{t=1}^N (w_j(t)-\hat w_j(t|t-1,\eta_j)^2
    + \lambda_j \cdot \|\eta_j\|_2 \right\}
    \label{eq:glasso1}
\end{equation}
with the one-step-ahead predictor \eqref{eq:pred_step2_MISO}, and $\eta_{j}$ being the vector of parameters related to the modules $G_{ji}$ for $i \in \mathcal{N}_j$, and related to the modules $\bar H_{js}$ for $s = 1,\dots, p$; $\lambda_j$ is the tuning parameter (penalization factor) of Glasso. The tuning of $\lambda_j$ is described in the numerical illustrations in Section \ref{sec:Num_ex}.\\
The right hand side of \eqref{eq:glasso1} is a mixed $l_1 /l_2$ norm. The Glasso estimate is a convex extension to lasso that penalizes groups of estimated parameters \cite{YuanGlasso}, imposing sparsity at group level. Within a group, it does not yield sparsity \cite{bach2011convex}.
If an appropriate penalization factor is chosen, only the dynamic modules that are actually present in the data generating network remain while the non-present terms are forced to $0$, thus providing an estimate of the structure of $\bar H$.

With either of the methods of Sections \ref{sec:323} or \ref{sec:324} the structure $\mathcal{V}_j$ of the disturbance model can be estimated entirely with convex and thus scalable methods, employing nonparametric (high-order ARX-) models. This structural information can be effectively used in the actual estimation of parametric dynamic models in the next Section.
\begin{remark}
It is possible to add regularization when estimating the high-order ARX models presented in this section to guarantee  stability of the estimates.
\end{remark}

\section{Estimating parametric network models}\label{sec:Id}
The next step in our identification procedure is
\begin{itemize}
    \item Step 3: Estimating a parametric network model.
\end{itemize}
While in Step 1 and 2 high-order (nonparametric) models \blue{of the same model order $n$} are used, and thus providing estimates with relatively high variance, in this step a parametric model is estimated from data where we exploit a very flexible Box-Jenkins model structure.
In Step 3 we extend the WNSF method \cite{Galrinho_Para_WNSF_2019}, and its application to dynamic networks in \cite{WNSF_stef}, to the reduced rank noise case such that we are able to obtain parametric models $G(\theta)$ and $H(\theta)$. \blue{The WNSF is in itself a three step method that starts with a high-order model before estimating the parametric model.}

\subsection{{Step 3.1: Refining the nonparametric model}}
By fixing the \blue{correctly} estimated disturbance topology obtained in the previous section we obtain consistent estimates of $\eta_j$ using one-step-ahead predictor \eqref{eq:pred_step2_MISO} \blue{defined in \eqref{eq:def_pred_varepsilon_a}, leading to a high-order ARX model with structured disturbance model.} The conditions for consistency of $\hat{\eta}^n_{j_N}$ are derived in Section \ref{sec:Analyses}.
\blue{By employing the structured disturbance model we reduce the variance of $\hat{\eta}^n_{j_N}$, while the model order $n$ remains the same.}\\
%
%
\blue{Using the consistent estimate $\etaNj$, we update the reconstructed innovation. Subsequently, we again update the high-order ARX model by replacing ${\varepsilon}_{a}(\hat{\zeta}^n_{j_N})$ with the updated reconstructed innovation ${\varepsilon}_{a}(\hat{\eta}^n_{j_N})$ in \eqref{eq:pred_step2_MISO}, and use this updated predictor to re-estimate $\eta_j$. This latter estimate can be seen as the starting high-order model for the WNSF method.}
%
%
At this point we \blue{still} have a high variance on the estimates of $\eta$ but negligible bias if model order $n$ \blue{throughout all the steps} is chosen sufficiently large. 
In the next step we reduce the variance by reducing the number of parameters to estimate\blue{, where we will make the step from a high-order (nonparametric) model to a parametric model}.

\subsection{{Step 3.2: Parametric model estimate}}

On the basis of the nonparametric model estimate characterized by $\hat\eta^n_{j_N}$ we are now going to estimate a parametric model of the dynamic  network by utilizing a Box Jenkins model structure:
\begin{equation}
\begin{aligned}
    G_{jl}(q,\theta) &= \frac{\quad \,\,\,\, l^{jl}_1q^{-1} + \dots + l^{jl}_{m_l} q^{-m_l}}{\,1+ f^{jl}_1q^{-1} + \dots + f^{jl}_{m_f} q^{-m_f}}, \\
        H_{jj}(q,\theta) &=  \frac{1+c^{jj}_1q^{-1} + \dots + c^{jj}_{m_c} q^{-m_c}}{\, 1+ d^{jj}_1q^{-1} + \dots + d^{jj}_{m_d} q^{-m_d}},\\
            H_{js}(q,\theta) &= \frac{\quad \,\,\,\, c^{js}_1q^{-1} + \dots + c^{js}_{m_c} q^{-m_c}}{\,1+ d^{js}_1q^{-1} + \dots + d^{js}_{m_d} q^{-m_d}}, \ s \neq j
\end{aligned}
\label{eq:G0H0_def_para}
\end{equation}
that can be rewritten as
\begin{equation}
\begin{aligned}
    G_{jl}(q,\theta)= \frac{L_{jl}(q,\theta)}{F_{jl}(q,\theta)} , \quad H_{js}(q,\theta)=\frac{C_{js}(q,\theta)}{D_{js}(q,\theta)} .
    \label{eq:relation_rational}
\end{aligned}
\end{equation}

From $G_{jl}(\etaNj)$ and $\bar H_{js}(\etaNj)$ that are obtained in \blue{the previous step} through the predictor \eqref{eq:pred_step2_MISO}, we can derive a related estimate of $H^0(q)$ according to \eqref{eq:Hbreve-I} leading to $H(\hat\eta^n_N) = \bar H(\hat\eta^n_N) + \begin{bmatrix} I \\ \Gamma(\hat\eta^n_N) \end{bmatrix}$, with $\Gamma(\hat\eta^n_N)$ an estimate of the direct feedthrough term $\Gamma^0$ of $H_b^0$, and that based on the relation $ \breve e_b(t) = \Gamma^0 \breve e_a(t)$ from \eqref{eq:def_H_Hbreve}, can be given by
\begin{equation}
    \Gamma(\hat\eta^n_N)\! =\! \Big( \frac{1}{N}\! \sum^N_{t=1}\! \varepsilon_b (\hat{\eta}^n_N)  \varepsilon_a^\top (\hat{\eta}^n_N)   \Big) \Big( \frac{1}{N}\! \sum^N_{t=1}\!   \varepsilon_a (\hat{\eta}^n_N)  \varepsilon_a^\top (\hat{\eta}^n_N)   \Big)^{-1}\!\!.
    \label{eq:Gamma_est}
\end{equation}
Following the WNSF approach, we are now going to fit the parametric Box Jenkins model to the nonparametric model estimated from Step 3.1, by solving for $\theta$ in the equations
\begin{equation}
\begin{aligned}
  &  F_{jl}(\theta) G_{jl}(\hat\eta^n_N) -L_{jl}(\theta)=0 \ , \\
   & D_{js}(\theta) H_{js}(\hat\eta^n_N) -C_{js}(\theta)=0.
   \label{eq:relation_rewritten_WNSF}
\end{aligned}
\end{equation}
However, since these equations can not be solved exactly, an optimization problem is formulated \cite{Galrinho_Para_WNSF_2019} that comes down to minimizing the quadratic residual vector on the equations \eqref{eq:relation_rewritten_WNSF} by solving (in node-wise notation):
\begin{equation}
    \min_{\theta_j} \|\etaNj - Q_j(\etaNj)\theta_j\|_2
    \label{eq:Q_wnsf}
\end{equation}
where
\begin{equation}
    Q_j(\eta)= \begin{bmatrix}
        Q_j^{g} & 0 \\ 0 & Q_j^{h}
    \end{bmatrix},
\end{equation}
with $Q_j^{g}$ and $Q_j^{h}$ diagonal matrices with entries
\begin{equation}
\begin{aligned}
  &  Q_j^{g^{jl}} (\eta)= \begin{bmatrix}
  -\mathcal{T}_{n \times m_f} [  G_{jl}(\eta) ] &
   \bar{I}_{n\times m_l}\end{bmatrix},\\
  &  Q_j^{h^{js}} (\eta)= \begin{bmatrix} -\mathcal{T}_{n \times m_d} [  H_{js}(\eta) ] &  \bar{I}_{n\times m_c} \end{bmatrix},
\end{aligned}
\end{equation}
\blue{with model orders $m_i, i \in \{l,f,c,d\}$ according to \eqref{eq:G0H0_def_para},} the top left corner of $\bar{I}_{n\times m}$ is $I_{m\times m}$ and has zeros otherwise, and $\mathcal{T}_{n\times m} [X_{ji}(q)]$ is a lower triangular Toeplitz matrix
where the first column is $\begin{bmatrix} x_0^{ji} & \cdots & x_{n-1}^{ji}  \end{bmatrix}^\top$ with $X_{ji}(q)= \sum^\infty_{k=0} x_k^{ji} q^{-k}$. \\
The problem \eqref{eq:Q_wnsf} is solved in first instance through the analytical least squares solution
\begin{equation}
    \hat{\theta}_{j_N}^{[0]} = \big(
    Q_j^\top (\hat{\eta}_{j_N}^n)
    Q_j (\hat{\eta}_{j_N}^n)
    \big)^{-1}
    Q_j^\top (\hat{\eta}^n_{j_N})
    \hat{\eta}_{j_N}^n.
    \label{eq:WNSF_theta0}
\end{equation}
However, a parameter estimate with smaller variance can be achieved if a weighted least squares criterion is applied\footnote{As an alternative we can consider a weighted least squares criterion to obtain $\hat{\theta}_{j_N}^{[0]}$ \eqref{eq:WNSF_theta0}, with the covariance matrix of the nonparametric model as weight.}. This is introduced in the next step.

\subsection{Step 3.3: Re-estimation of parametric model}

In this step we reduce the variance further by re-estimating the obtained parametric models $G(\theta)$ and $H(\theta)$ defined in \eqref{eq:relation_rational}.
For a statistical optimal solution of \eqref{eq:Q_wnsf}, instead of the standard least squares problem \eqref{eq:Q_wnsf}, a weighted least squares problem should be solved, where the optimal weight is given by the inverse of the covariance matrix of the residual $\etaNj - Q_j(\etaNj)\theta_j^0$, with $\theta_j^0$ the actual network coefficients related to node $w_j$. This is not directly applicable since $\theta_j^0$ is unknown. However it can be shown \cite{Galrinho_Para_WNSF_2019} that
\begin{equation}
   \etaNj - Q_j(\etaNj) \theta_j^0 = T_j(\theta^0_j) (\etaNj - \eta^{n0}_{j} )
    \label{eq:WNSF_T_error},
\end{equation}
with $\eta_j^{n0}$ the real network coefficients related to the $\eta$-parametrized ARX model
and $T_j(\theta)$ a block diagonal matrix with the denominator polynomials as entries
\begin{equation}
\begin{aligned}
   & T_j^{g^{jl}}(\theta) = \mathcal{T}_{n\times n} [F_{jl}(\theta)],\\
   & T_j^{h^{js}}(\theta) = \mathcal{T}_{n\times n} [D_{js}(\theta)],
\end{aligned}
\end{equation}
where $\mathcal{T}_{n\times n} [X_{ji}(q)]$ is a lower triangular Toeplitz matrix
where the first column is $\begin{bmatrix} 1 & x_1^{ji} & \cdots & x_{m}^{ji} & 0_{n-m-1}  \end{bmatrix}^\top$ with $X_{ji}(q)= 1+\sum^\infty_{k=1} x_k^{ji} q^{-k}$.

Result \eqref{eq:WNSF_T_error} motivates the use of a weighted least estimator with weighting matrix
\[
W_j = T_j^{-1}(\theta_j^0) (P_{\etaNj})^{-1} T_j^{-T}(\theta_j^0)
\]
with $P_{\etaNj}$ the covariance matrix of the nonparametric model. This
can be implemented in an iterative scheme according to
\begin{equation}
\begin{aligned}
    &\hat{\theta}_{j_N}^{[k+1]} =\\
    &\big(
    Q_j^\top (\hat{\eta}^n_{j_N})
    W_j (\hat{\theta}^{[k]}_{j_N})
    Q_j (\hat{\eta}^n_{j_N})
    \big)^{-1}
    Q_j^\top (\hat{\eta}^n_{j_N})
    W_j (\hat{\theta}^{[k]}_{j_N})
    \hat{\eta}^n_{j_N}.
    \label{eq:WNSF_theta_end}
    \end{aligned}
\end{equation}
For consistency of the estimates of parameter vector $\theta$ we refer to the proof in the WNSF method \cite{Galrinho_Para_WNSF_2019}, with the actual model orders $m_i$ with $i = f, l, c, d$ \eqref{eq:G0H0_def_para} known.
{\remark{Because in this final step we correct for the variance due to the modeling error \eqref{eq:WNSF_T_error}, the final estimate will have a reduced variance.}}

Throughout the presented steps we split the MIMO optimization into $L$ \blue{linear regressions} that rely on explicit analytical solutions, and that allows for parallel computing. The Algorithm is given as follows.

\begin{Algorithm}
{Algorithm for full network identification in dynamic networks, including disturbance topology detection}\\
\vspace{-5pt}
\hrule
\vspace{-3pt}
\textbf{Inputs:} $w(t),r(t)$\blue{, $R^0(q)$, model orders $m_i, i \in \{l,f,c,d\}$, network topology}.\\
\textbf{Output:} Disturbance topology, $\hat{\theta}_N$.\\

\vspace{-20pt}
    Disturbance topology detection
        \begin{enumerate}[label*=\arabic*.] 
            \item Estimate noise rank $p$ based on the reconstructed innovation ${\varepsilon}(t,\hat{\zeta}^n_N)$ \eqref{eq:innovation_1}, and if $p<L$ order the nodes.
            \item
            \begin{enumerate}[label*=\arabic*]
                \item Obtain consistent estimate $\hat{\zeta}^n_N$ with least squares solution \eqref{eq:zetax2}, where the nodes are ordered and by utilizing the estimated noise rank $p$.
                \item
            Use the reconstructed innovation ${\varepsilon}_a(t,\hat{\zeta}^n_N)$ as measured input in the one-step-ahead predictor \eqref{eq:pred_step2_MISO} defined in \eqref{eq:def_pred_varepsilon_a} to estimate the noise correlation structure. We use
            \begin{enumerate}
                \item Structure selection with AIC, BIC and CV,
                \item Glasso,
            \end{enumerate}
            applied to estimate $\hat{\eta}^n_{j_N}$ that is obtained with least squares solution \eqref{eq:etax}. 
        \end{enumerate}
        \end{enumerate}

     Estimating parametric network models
        \begin{enumerate}[label*=\arabic*.]
        \setcounter{enumi}{2}
            \item   \begin{enumerate}[label*=\arabic*]
            \item Refine the nonparametric ARX model and obtain consistent estimate $\hat{\eta}^n_N$ with one-step-ahead predictor \eqref{eq:pred_step2_MISO}, where the estimated disturbance topology is fixed and update the reconstructed innovation
            %
            to ${\varepsilon}_a(t,\hat{\eta}^n_N)$ to re-estimate $\hat{\eta}^n_N$.
                \item Reduce the nonparametric ARX model to a parametric model and obtain initial estimate $\hat{\theta}_N^{[0]}$ by \eqref{eq:WNSF_theta0}.
                \item Re-estimate $\hat{\theta}_{j_N}^{[k+1]}$ with \eqref{eq:WNSF_theta_end}, where we update the weighting matrix $W_j (\hat{\theta}^{[k]}_{j_N})$ in each iteration.
                %
            \end{enumerate}
        \end{enumerate}
 \vspace{5pt}
\hrule
\end{Algorithm}
We continue to iterate until we have reached the convergence criterion $    \tfrac{\| \hat{\theta}_N^{[k]} -\hat{\theta}_N^{[k-1]} \|}{\| \hat{\theta}_N^{[k-1]} \|} <0.0001$. This convergence criterion is also used in the simulation results in Section \ref{sec:Num_ex}.
In the next Section we derive the conditions required for consistency of estimates $\hat{\zeta}^n_{j_N}$ and $\hat{\eta}^n_{j_N}$.
\section{Theoretical analyses}\label{sec:Analyses}
%
From here on we consider $n=n(N)$ i.e. the model order $n$ increases as the data length $N$ increases, \blue{while with increasing $N$, $n/N$ tends to $0$ with a particular rate} \cite{Wahlberg_Ljung_As_Prop,Galrinho_Para_WNSF_2019}.\\
%
 Next we derive the conditions under which the estimates $\hat{\zeta}^n_N$ and $\hat{\eta}^n_N$, and consequently the reconstructed innovation are consistent.
\subsection{Consistency of $\hat{\zeta}^n_N$ in Step 2.1: Refining the nonparametric model}\label{sec:anal1}
With the noise rank $p$ available and the nodes ordered we gained structural information on the unique noise model $\breve{H}^0(q)$ \eqref{eq:def_H_Hbreve}, namely we know that for the reduced noise rank case $p<L$ the last $L-p$ columns in $\breve{H}^0(q)$ are $\begin{bmatrix} 0 & I
\end{bmatrix}^\top$. Moreover, taking the inverse of $\breve{H}^0 (q)$ does not affect the last $L-p$ columns since
\begin{equation}
    (\breve{H}^0)^{-1} =
    \begin{bmatrix}
    (H_a^0)^{-1} & 0 \\
    -\big(  H_b^0 - \Gamma^0   \big) (H_a^0)^{-1} & I
    \end{bmatrix}.
    \label{eq:h_breve_inverse}
\end{equation}
%
%
%
As a result the term $(\breve{H}^0(q))^{-1} R^0(q)$ in the one-step predictor \eqref{eq:x1}, has the following structure
\begin{equation}
    (\breve{H}^0)^{-1} R^0 =  \begin{bmatrix}
       (H_a^0)^{-1} R^0_a & 0 \\
       -\big(  H^0_b - \Gamma^0   \big) (H^0_a)^{-1}  R^0_a& R^0_b
    \end{bmatrix},
\end{equation}
with the second block column consisting of known terms only. This allows in the parametrization of the predictor \eqref{eq:predictorARX1x} to replace the square polynomial $\breve B(\zeta)$ with a non-square polynomial $B(\zeta)$, leading to
\begin{equation}
\begin{aligned}
    \hat{w} (t|t-1,\zeta) & =\big(I- \breve A(\zeta)\big) w(t) + B(\zeta) r_a(t) +
    \begin{bmatrix}
        0\\R_b^0
    \end{bmatrix} r_b(t) \\
    & =\varphi(t)\zeta + \begin{bmatrix}
        0\\R_b^0
    \end{bmatrix} r_b(t)
    \label{eq:pred_step1_rr_0},
\end{aligned}
\end{equation}
with $\varphi(t)$ composed of the appropriate terms in $w$ and $r_a$.\\
Note that for an actual network with representation $G^0, \breve H^0, R^0$, the one-step predictor is still given by \eqref{eq:x1}, but now the predictor model \eqref{eq:pred_step1_rr_0} can use the known external excitation signals $r_b(t)$. The ARX model is estimated according to $\hat \zeta^n_N = \arg\min_{\zeta} \frac{1}{N} \sum_{t=1}^N \varepsilon^T(t,\zeta)\varepsilon(t,\zeta)$, with $\varepsilon(t,\theta) = w(t) - \hat w(t|t-1;\zeta)$, leading to the analytical solution:
\begin{equation}
  \label{eq:zetax2}
   \hat{\zeta}^n_{N}=
    \Bigg[ \!\frac{1}{N}\! \sum^N_{t=1}\! \varphi(t) \varphi^{\top}\! (t) \Bigg] ^{-1}
     \!\! \frac{1}{N}\! \sum^N_{t=1}\! \varphi(t)\!  \left[ w(t)\! -\! \begin{bmatrix} 0\\R_b^0 \end{bmatrix} r_b(t) \right].
\end{equation}
Note that Remark 1 holds and therefore predictor \eqref{eq:pred_step1_rr_0} can be decomposed in separate \blue{predictors for each node}.
The conditions for consistency are formulated in Proposition \ref{prop:1} and the proof is added in the appendix.
\begin{proposition}\label{prop:1}
Consistency $\hat{\zeta}^n_N$\\
Consider a dynamic network that satisfies Assumption 1.
Additionally, consider the one-step-ahead predictor \eqref{eq:pred_step1_rr_0}.
Then the transfer function matrices $(\breve{H}^0(q))^{-1} (I-G^0(q))$ and $(\breve{H}^0(q))^{-1} \begin{bmatrix} R_a^0(q)^\top& 0\end{bmatrix}^\top$
are consistently estimated with the analytical solution \eqref{eq:zetax},
if the following conditions hold:
\begin{enumerate}
    \item The external excitation $r(t)$ is uncorrelated to the noise $e(t)$.
    \item {The spectral density of $\kappa(t) = \begin{bmatrix} r_a(t)^\top & w(t)^\top \end{bmatrix}^\top$, $\Phi_\kappa(\omega)>0$ for a sufficiently high number of frequencies $\omega$.}
    \item $\breve A(q,\zeta)$ and $ B(q,\zeta)$ are of high order, such that $n\to \infty$.
    %
\end{enumerate}
\end{proposition}
\begin{remark}\label{col:1}
Condition (1) and (2) of Proposition 1 are given for all signals present in the network.
These conditions remain unchanged when we convert from a MIMO predictor to $L$ \blue{linear regressions}. Therefore the proof also holds for a predictor assessed per node.
\end{remark}
{\textbf{Proof}}: See appendix.

\subsection{Consistency of $\hat{\eta}^n_N$ in Step 3.1: Refining the nonparametric model}
A refined nonparametric model is estimated by exploiting the information on the noise topology in the form of a structured polynomial model $B(\eta_j)$ for $\bar H_{js}(\eta_j)$ in the predictor \eqref{eq:pred_step2_MISO}, leading to the analytical solution 
\begin{equation}
  \label{eq:etax}
   \hat{\eta}^n_{N}=
    \Bigg[ \!\frac{1}{N}\! \sum^N_{t=1}\! \varphi(t) \varphi^{\top}\! (t) \Bigg] ^{-1}
     \!\! \frac{1}{N}\! \sum^N_{t=1}\! \varphi(t)\!  \left[ w(t)\! -\! R^0 r(t) \right].
\end{equation}
with $\varphi(t)$ composed of the appropriate terms in $w$ and $\varepsilon(\hat{\eta}^n_N)$.\\
%
%
The conditions for consistency are formulated in Proposition \ref{prop:2}.
\begin{proposition}\label{prop:2} Consistency {$\hat{\eta}^n_N$}\\
Consider a dynamic network that satisfies Assumption 1 and Proposition \ref{prop:1}, and assume the disturbance topology is estimated correctly.
Additionally, consider the one-step-ahead predictor \eqref{eq:pred_step2_MISO} for all $j$. Then the transfer function matrices of $G^0(q)$ and $\breve H^0(q)-I$ are consistently estimated with the analytical solution $\hat{\eta}^n_N$ \eqref{eq:etax},
if the following conditions hold:
 \begin{enumerate}
    \item For all $j$, the spectral density $\Phi_{\bar{\kappa}}(\omega)$  of $\bar{\kappa}(t) := \begin{bmatrix} w_{\{ \mathcal{N}_j\}}(t)^\top & e_{\{ \mathcal{V}_j\}}(t) ^\top \end{bmatrix}^\top$, satisfies $\Phi_{\bar{\kappa}}(\omega)>0$ for a sufficiently high number of frequencies $\omega$.
    \item \label{cond2} The data generating system is in the model set, i.e. there exists a $\eta_0$ such that {$G(q,\eta_0)=G^0(q)$ and $\bar{H}(q,\eta_0)= \breve H^0(q)-I$.}
\end{enumerate}
\end{proposition}
\textbf{Proof}: See appendix.

With consistent estimate $\hat{\eta}^n_N$ we can update the reconstructed innovation ${\varepsilon}(t,\hat{\eta}^n_N)=\begin{bmatrix}
   {\varepsilon}_a(t,\hat{\eta}^n_N)^\top & {\varepsilon}_b(t,\hat{\eta}^n_N)^\top
\end{bmatrix}^\top$ consistently for each time step $t=1,\dots,N$
\begin{equation}
 \begin{aligned}
    {\varepsilon}(t,\hat{\eta}^n_N) &\to \breve{e}(t) \qquad \text{w.p. 1 as} \, N\to \infty \, \forall t ,\\
\end{aligned}
\end{equation}
where the innovation is reconstructed per node according to $\varepsilon_j(t,\eta)= w_j(t) -  \hat{w}_j(t|t-1,\eta)$ using one-step-ahead predictor \eqref{eq:pred_step2_MISO}.

\begin{remark}\blue{Note that Condition \ref{cond2} of Proposition \ref{prop:2} incorporates the condition that the noise rank $p$ is chosen correctly, and the disturbance model is flexible enough to represent the exact disturbance topology of the network.}
\end{remark}

%
Following the line of reasoning in \cite{Hof2020path}, the spectral conditions in Propositions \ref{prop:1} and \ref{prop:2}, which are actually data informativity conditions, can generically be replaced by path-based conditions on the graph of the network model set.

\subsection{Generic data informativity conditions}
Condition (2) of Proposition \ref{prop:1} and Condition (1) of Proposition\ref{prop:2} is a spectral data informativity condition on internal node signals in $w$, and it is difficult to interpret it for an experimenter. In this section we replace the spectral condition with a path-based data informativity condition in a generic sense\footnote{\blue{Genericity is considered in the sense that the corresponding property holds for almost all models in the model set, possibly excluding a set of measure $0$.}}, i.e. independent of the numerical values of the network dynamics.
By doing so we can evaluate if data informativity is satisfied based on the network and disturbance topology, and the properties of the external signals.
Next we formulate the conditions in terms of properties and locations of the external signals analogous to Lemma 1 and Proposition 1 from \cite{Hof2020path}, by means of vertex-disjoint paths from external signals to internal node signals\blue{, where two paths are vertex-disjoint if they have no nodes in common, including their start and end nodes \cite{Woude1991graph}}. The consequences are illustrated in a 6-node example.
\subsubsection{Vertex-disjoint paths}
%
%
The generic version of Condition (2) of Proposition \ref{prop:1} is given in Proposition \ref{prop:3}.
\begin{proposition}\label{prop:3}
%
The spectrum condition $\Phi_\kappa(\omega)>0$ for $\kappa(t) = \begin{bmatrix} r_a(t)^\top & w(t)^\top \end{bmatrix}^\top$  in Condition (2) of Proposition \ref{prop:1} is generically satisfied if there are $L$ vertex-disjoint paths from $\begin{bmatrix} r_b(t)^\top & e(t)^\top \end {bmatrix}^\top$ to $w(t)$.
\end{proposition}
{\textbf{Proof:}} See appendix.

Proposition \ref{prop:3} gives a sufficient generic path-based condition that requires to have external excitation signals at certain locations in the network, combining data informativity conditions with identifiability \cite{Hof2020path}.

The set $\mathcal{V}$ denotes the set of indices of all the disturbing noise signals, where $\mathcal{V}_j$ is a subset of $\mathcal{V}$.
For the generic condition for Condition (1) of Proposition \ref{prop:2} we introduce notation $e_{\{ \mathcal{X}_j\}}(t)$, where $\mathcal{X}_j$ is the set of indices of all the disturbing noise signals excluding indices that are already present in set $\mathcal{V}_j$, i.e. $\mathcal{X}_j = \mathcal{V}{/\mathcal{V}_j}$. 
%
%
%
\begin{proposition}\label{prop:4}
The spectrum condition $\Phi_{\bar{\kappa}}(\omega)>0$ for $\bar{\kappa}(t) = \begin{bmatrix} w_{\{ \mathcal{N}_j\}}(t)^\top & e_{\{ \mathcal{V}_j\}}(t)^\top \end{bmatrix}^\top$  in Condition (1) of Proposition \ref{prop:2} is generically satisfied if
there are Cardinal$\{ \mathcal{N}_j\}$ vertex-disjoint paths from $\begin{bmatrix}r(t)^\top & e_{\{ \mathcal{X}_j\}}(t)^\top \end{bmatrix}^\top$ to $w_{\{ \mathcal{N}_j\}}(t)$.
\end{proposition}
{\textbf{Proof:}} See appendix

Proposition \ref{prop:4} gives a sufficient generic path based condition that requires external excitation signals at certain locations such that $\Phi_{\bar{\kappa}}(\omega)>0$ for a sufficiently high number of frequencies.
\begin{remark}
If we want to identify only the $j^{th}$ row of the network (or only part of the network), we can consider the predictor in Proposition \ref{prop:2} only for node $j$ and satisfy the conditions in Proposition \ref{prop:2} and \ref{prop:4} for node $j$.
\end{remark}
Next we elaborate the vertex-disjoint path conditions by means of an example where a network is subject to reduced rank noise.
\subsubsection{Reduced rank noise example}\label{subsec:example_rr}
\begin{figure}[!hb]
\begin{center}
\includegraphics[width=8.5cm]{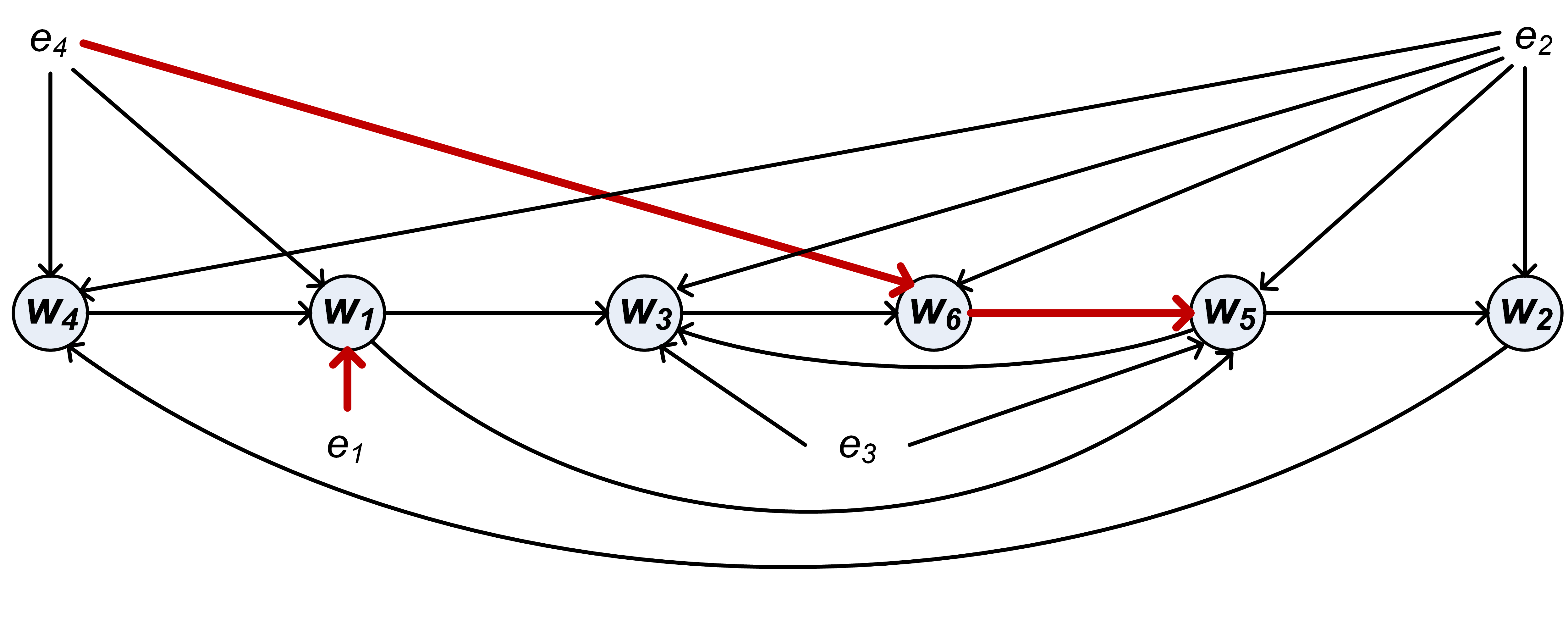}
\caption{6-node dynamic network with reduced rank noise that has rank $p= 4$, \blue{no $r(t)$ signals are shown. 
The arrows represent the edges for which $G^0_{ji}\neq 0$ and $H^0_{ji}\neq0$, where the arrows indicated in red are examples of the two vertex disjoint paths needed to satisfy Proposition 4 for output $w_3(t)$}} 
\label{fig:6node_fig}
\end{center}
\end{figure}

We consider a 6-node network that satisfies Assumption 1 and is subject to reduced rank noise of rank $p=4$ shown in Figure \ref{fig:6node_fig}. This 6-node example is additionally used in the simulations in Section \ref{sec:Num_ex}, and is further defined in Appendix \ref{App:used_models}. The nodes are ordered such that the first $p$ nodes are subject to full rank noise. Moreover, we assume the disturbance topology is correctly estimated.

The goal of this example is to elaborate on the path-based data informativity conditions given in Proposition \ref{prop:3} and \ref{prop:4}.
To be more specific, we show which external excitation signals are sufficient in order to satisfy the spectral Condition (2) in Proposition \ref{prop:1} and Condition (1) in Proposition \ref{prop:2}.
In the example we have external noise signals $e(t)=\begin{bmatrix}    e_1(t) &\dots &e_4(t)\end{bmatrix}^\top$ and external excitation signals $r_k(t)$, for simplicity we assume $R^0$ contains elements that are either 0 or 1.
%

In order to satisfy Proposition \ref{prop:3}, we require $L=6$ vertex-disjoint paths from $\begin{bmatrix} r_b(t)^\top & e(t)^\top \end{bmatrix}^\top$ to $w(t)$. The first $p=4$ nodes, denoted by $w_a(t)$, are excited by the noise $e(t)$; we therefore require at least $L-p=2$ external excitation signals $r_k(t)$ on the last 2 nodes $w_b(t)=\begin{bmatrix}
    w_5(t)& w_6(t)
\end{bmatrix}^\top$, i.e $r_b (t) = \begin{bmatrix}
  r_5(t)& r_6(t)
\end{bmatrix}^\top$ with $R_b=I\in\mathbb{R}^{2\times2}$. Therefore we satisfy Proposition \ref{prop:3} since we have 6 vertex-disjoint paths from $\begin{bmatrix}
    e(t)^\top & r_b(t)^\top
\end{bmatrix}^\top$ to $ \begin{bmatrix}
    w_a(t)^\top &w_b(t)^\top
\end{bmatrix}^\top$.

\blue{To show how Proposition 4 is satisfied, we first consider output node $w_3(t) = G_{31}(\eta)w_1(t)+ G_{35}(\eta)w_5(t)+ H_{32}(\eta)e_2(t)+H_{33}(\eta)e_3(t)$, that has $w_{\{\mathcal{N}_3\}}(t)= \begin{bmatrix}w_1(t) & w_5(t) \end{bmatrix}^\top$ and $e_{\{\mathcal{V}_3\}}(t)= \begin{bmatrix}e_2(t) & e_3(t) \end{bmatrix}^\top$. We need Cardinal$\{ \mathcal{N}_3\}=2$ vertex-disjoint paths from $\begin{bmatrix}r(t)^\top & e_{\{ \mathcal{X}_j\}}(t)^\top \end{bmatrix}^\top$ to $w_{\{ \mathcal{N}_3\}}(t)$. There already exist 2 vertex disjoint paths from $e_{\{ \mathcal{X}_j\}}(t)= \begin{bmatrix}e_1(t) & e_4(t) \end{bmatrix}^\top$ to $w_{\{\mathcal{N}_3\}}(t)$. This shows that Proposition \ref{prop:4} is satisfied by the two vertex disjoint paths from $e_1 (t)\rightarrow w_1(t)$ and from $e_4(t)\rightarrow w_6(t) \rightarrow w_5(t)$ as indicated in red in Figure \ref{fig:6node_fig}.}
If we apply the same reasoning to the other nodes we see that for node
\begin{itemize}
    \item $w_1(t)$ with $w_{\{\mathcal{N}_1\}}(t)= w_{4}(t)$, there exists a vertex-disjoint path from $e_2(t) \rightarrow w_4(t)$.
    \item $w_2(t)$ with $w_{\{\mathcal{N}_2\}}(t)= w_{5}(t)$, there exists a vertex-disjoint path from $e_3(t) \rightarrow w_5(t)$.
    \item $w_4(t)$ with $w_{\{\mathcal{N}_4\}}(t)= w_2(t)$, there exists a vertex-disjoint path from $e_3(t)\rightarrow w_5(t) \rightarrow w_2(t)$ 
    \item $w_5(t)$ with $w_{\{\mathcal{N}_5\}}(t)= \blue{ \begin{bmatrix}w_1(t) & w_6(t) \end{bmatrix}^\top}$, there exist 2 vertex-disjoint paths from $e_1(t)\rightarrow w_1(t)$ and from $e_4(t)\rightarrow w_6(t)$.
    \item $w_6(t)$ with $w_{\{\mathcal{N}_3\}}(t)= w_3(t)$, there exists a vertex-disjoint path from $e_3(t)\rightarrow w_3(t)$.
\end{itemize}
In order to satisfy Proposition \ref{prop:4} we therefore do not require additional external excitation signals $r_k(t)$. \\
Consequently, in order to identify the full network for the given example, it is sufficient to add external signals $r_b(t)=\begin{bmatrix}
    r_5(t) & r_6(t)
\end{bmatrix}^\top$
with $R_b=I\in \mathbb{R}^{2\times 2}$
 that satisfies Proposition \ref{prop:3}.
\section{Numerical simulations}\label{sec:Num_ex}
In this section we show the results of different steps in Algorithm 1. We assume $R^0=I$, and consider the system given in Figure \ref{fig:6node_fig} and Appendix \ref{App:used_models}.

For the simulation study we use normally distributed zero mean white external signals, where $\{r(t)\}$ has a variance of $5$ and the vector of $e$-signals has variances $\{ 0.1,\, 0.2,\, 0.3,\, 0.4 \}$. We simulate the nodes according to $w(t) = ( I-G^0 )^{-1} ( R^0 r(t) + H^0 e(t) )$ and perform $M=100$ Monte Carlo runs over five data lengths logarithmically spaced between $300$ and $50000$. \blue{For each of the data lengths $N$ a specific value of the model order $n$ is chosen according to $n =  10,\, 20 , \,30 ,\, 40,\, 40$, for increasing values of $N$. The actual model orders $m_i, i \in \{l,f,c,d\}$ can be derived from Appendix \ref{App:used_models}.}

Next we describe the noise rank estimation results of step 1 of Algorithm 1.
\subsection{Rank $p$ and ordering of the nodes}
In order to obtain the noise rank $p$ we perform a rank test (singular value decomposition) on covariance matrix $\hat{\Lambda}$ \eqref{eq:lambda_tildex}.
For data length $N=300$, the singular values averaged over the 100 Monte Carlo runs are $\blue{svd(\hat \Lambda_N)} \!=\! \begin{bmatrix}
    0.37 & 0.26 & 0.21 & 0.06 & 2.13\! \cdot \! 10^{-8} & 1.96\! \cdot \! 10^{-9}
\end{bmatrix}$, where we see that the last two singular values are close to zero. As data length increases the last two values converge even closer to zero. For $N=50000$ we obtain the following averaged singular values $\blue{svd(\hat \Lambda_N)}\! =\! \begin{bmatrix}
    0.59 & 0.40 & 0.39 & 0.10 & 4.04\! \cdot 10^{-13}\! & 1.24\! \cdot\!  10^{-13}
\end{bmatrix}$, 
\blue{showing that a clear gap between the fourth and fifth singular value points to a correct rank estimate of $4$.}
%

Finally with the noise rank $p$ available we can reorder the nodes such that $\begin{bmatrix} I_p & 0\end{bmatrix} \Pi^\top \hat{\Lambda} \Pi\begin{bmatrix} I_p & 0\end{bmatrix}^\top$ has rank $p$.
%

Next we show the disturbance topology detection results of step 2 of Algorithm 1.
\subsection{Topology estimation of the disturbance model}
For the topology detection we are interested in which indices belong in set $\mathcal{V}_j$ for all $j$, where the indices indicate where the edges are located in the disturbance model. We evaluate the performance of the topology detection by evaluating the trade-off between overestimating and underestimating the number of edges, that is typically used in receiver operating
characteristic (ROC) curves \cite{hajian2013ROC}.

If an edge is present in both the data generating disturbance and the estimated disturbance topology, we count this edge as a true positive (TP). If an edge is present in the estimated disturbance topology but does not exist in the data generating system, we count this edge as a false positive (FP). Additionally we let $Pos$ indicate the total number of existing edges and $Neg$ indicates the total number of non-existing edges in the disturbance model.
The ROC curve plots the true positive rate (TPR) versus the false positive rate (FPR), with
\begin{equation}
  TPR = \frac{TP}{Pos}  , \quad   FPR= \frac{FP}{Neg},
\end{equation}
where {FPR=0 and TPR=1 represented by the point $(0,1)$}, indicates the topology is perfectly reconstructed. We evaluate the closeness to the point $(0,1)$ by utilizing the distance function
\begin{equation}
    dis= \sqrt{ FPR^2+ (1-TPR)^2},
\end{equation}
%
%
  %
  \begin{figure}[b]
    \includegraphics[width=0.486\textwidth]{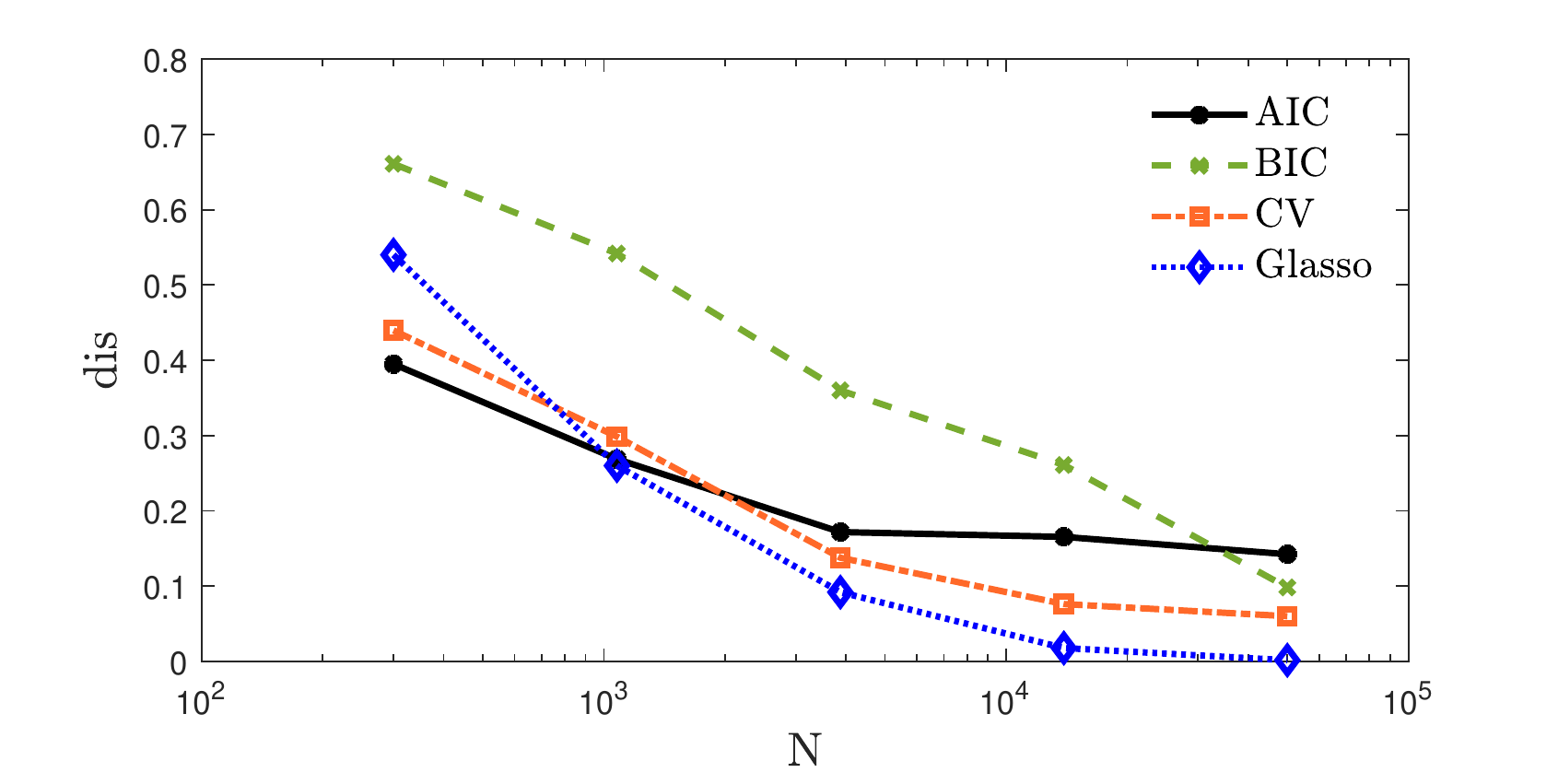}
  \caption{$dis$ as a function of $N$, averaged over the Monte Carlo runs.}
  \label{fig:dis_rr}
\end{figure}
For the structure selection procedure we test all possible combinations in set $\mathcal{V}_j$ and employ AIC, BIC and CV. For AIC we use
\begin{equation}
    \frac{1}{2} \text{log} \Big( V_{j_N}(\hat{\eta}^n_{j_N}) \Big) + \frac{n_{p_j}}{N},
\end{equation}
with $n_{p_j}$ the number of estimated parameters for node $j$ and
\begin{equation}
    V_{j_N}(\hat{\eta}^n_{j_N})= \frac{1}{N} \sum^N_{t=1} \varepsilon_j(t,\hat{\eta}^n_{j_N}  )^2.
\end{equation}
For BIC we use
\begin{equation}
    N*\text{log} \Big(  V_{j_N}(\hat{\eta}^n_{j_N})  \Big)+ N( \text{log}(2\pi)+1)+ n_{p_j} \text{log}(N).
\end{equation}
From these simulations we select set $\mathcal{V}_j$ that gives the smallest AIC or BIC value.
For the CV we split the data $Z^N= Z^{(1)}Z^{(2)}$ in a training set $Z^{(1)}$ of length $ \frac{2}{3}(N+1)$ and obtain the estimates for the different combinations in set $\mathcal{V}_j$ according to
\begin{equation}
\begin{aligned}
    \hat{\eta}_{j_N}^{(1)} &=  \underset{\eta }{\text{argmin}} V_{j_N}(\eta_j,Z^{(1)}),
\end{aligned}
\end{equation}
With the validation set $Z^{(2)}$, that contains the remaining data of length $N^{(2)}=\frac{1}{3}(N+1)$, we minimize objective function
\begin{equation}
\begin{aligned}
    V_{j_N}(\hat{\eta}_{j_N}^{(1)},Z^{(2)}) &= \frac{1}{N^{(2)}} \sum^{N^{(2)}}_{t=1} \varepsilon_j(t,\hat{\eta}_{j_N}^{(1)}  )^2,
\end{aligned}
\end{equation}
and select the set $\mathcal{V}_j$ that gives the smallest root mean squared error (RMSE)
\begin{equation}
    RMSE_j= \sqrt{ V_{j_N}(\hat{\eta}_{j_N}^{(1)},Z^{(2)}) }.
\end{equation}
For Glasso we fully parametrize the disturbance model, using the known topology of $G^0$ and fixed $R^0=I$.
%
We inspect all elements of the disturbance model matrix that is parametrized with the Glasso estimates \eqref{eq:glasso1}. If element $H_{ji}(\hat{\eta}_N) $ of the disturbance model matrix contains nonzero Glasso estimates we say this element contains dynamics, and therefore an edge is present and $i\in \mathcal{V}_j$. 
To prevent arbitrary small Glasso estimates are seen as dynamics we define a tolerance, where the Glasso estimates are nonzero if the $l_2$ norm of these estimates is larger than $10^{-3}$.
The choice to include the estimates of $G_{jl}(\eta)$ in the penalization is due to the implementation of Glasso \cite{boyd2011Glasso}. %
%
For good estimates on the disturbance topology, we utilize the known topology of $G^0$ and deal with known $R^0 r(t)$ signals appropriately.

Tuning of $\lambda_j$ is done via a grid based search similar to the CV structure selection. First we select a grid $\lambda_j^{grid} = \{ 0, 25, 50 , \cdots , 2000\}$ containing $\lambda_j$ values to test. For each grid point we estimate $\hat{\eta}_j^{grid}$ using Glasso, from where the topology is derived by inspecting the disturbance model for dynamics as mentioned before, and fix the topology $H^{grid}_j$ per node. Next we apply CV using topology $H^{grid}_j$ and estimate the $\text{RMSE}_j$. The grid point with the lowest $\text{RMSE}_j$ is selected as the $\lambda_j$ value. Repeating the tuning procedure over a number of runs gives the minimally required value for $\lambda_j$. The tuning procedure is applied to all nodes for the different data lengths $N$.

Figure \ref{fig:dis_rr} shows the topology detection results, with the distance averaged over 100 Monte Carlo runs. \blue{The BIC is a consistent information criterion \cite{schwarz1978estimating,Bayes_factors}, meaning that the estimated disturbance topology will converge to the actual topology if $N \to \infty$. However, as can be seen in the results in Figure \ref{fig:dis_rr}, the full convergence of the BIC procedure is not reached for the given data lengths. Until the BIC procedure converges to the actual disturbance topology, it tends to underestimate the number of edges
that are actually present, therefore the mismatch in the distance function is caused by not detecting all the TP's. The AIC is not a consistent information criterion, but has a faster convergence rate compared to the BIC \cite{zhang1993convergence_AIC_BIC}. The AIC tends to overestimate the number of edges, meaning the mismatch is caused by detecting the FP's. The CV is comparable to AIC but has a slower convergence rate. Finally the Glasso seems to have the best of both AIC and BIC. However, these results heavily depend on the selected tuning parameter $\lambda$, where it is not guaranteed that a suitable $\lambda$ exists.}


Next we show the parametric estimation results of step 3 of Algorithm 1\blue{, where we fix the estimated disturbance topology}. \blue{Based on the results in Figure \ref{fig:dis_rr} we have fixed the correctly estimated disturbance topology obtained with Glasso for $N=50000$, where $TPR=1$ and $FPR=0$.}
%
%
%
%
%
\subsection{Estimating the parametric model}
\begin{figure}[b]
    \includegraphics[width=8.7cm]{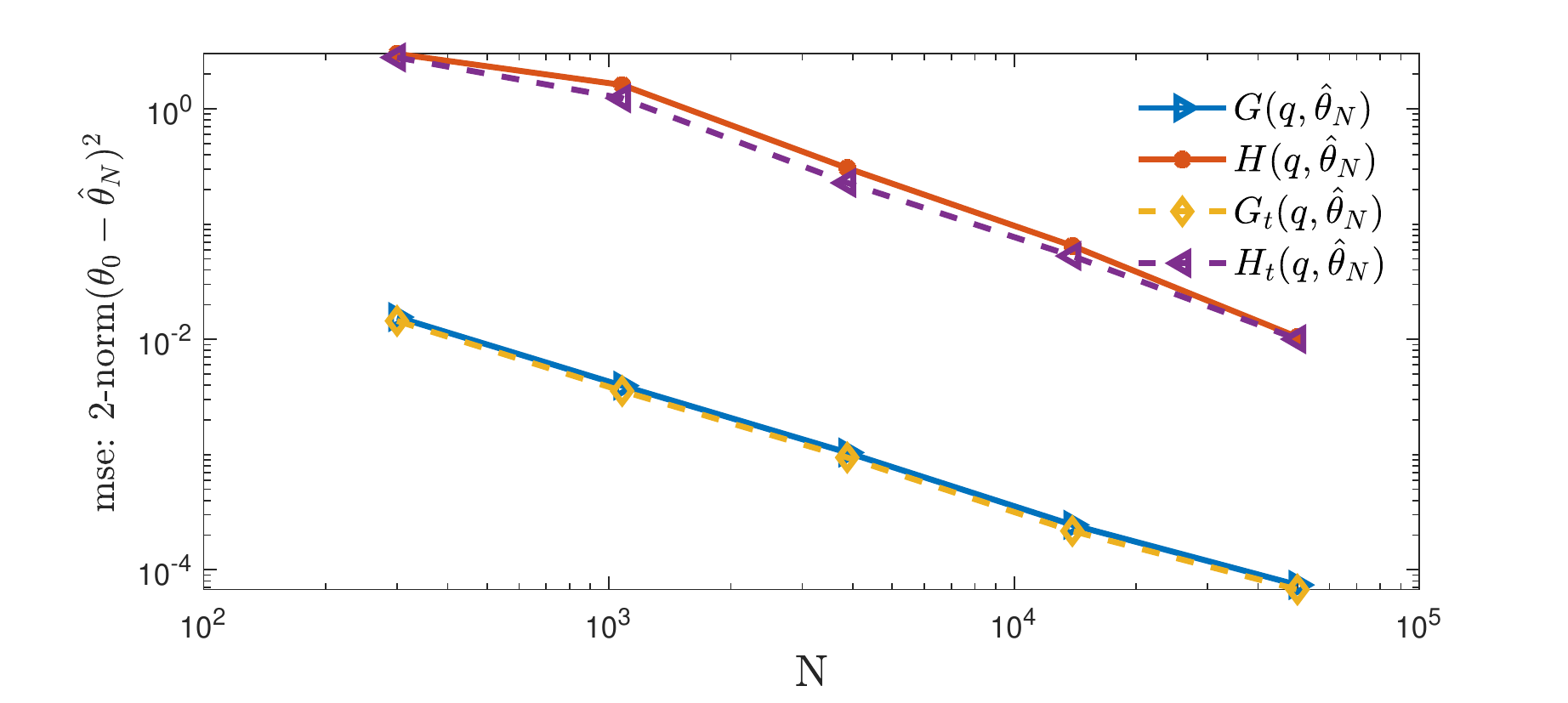}
  \caption{MSE between $\hat{\theta}_N$ and $\theta_0$ as function of sample size, averaged over the Monte Carlo runs, obtained with Algorithm 1 with $R^0=I$, where subscript $\{t\}$ indicates the use of the true (unknown) white noise as a predictor input instead of the reconstructed innovation.}
    \label{fig:reduced_rank_sim_MSE}
\end{figure}
\begin{figure}[b]
    \includegraphics[width=8.7cm]{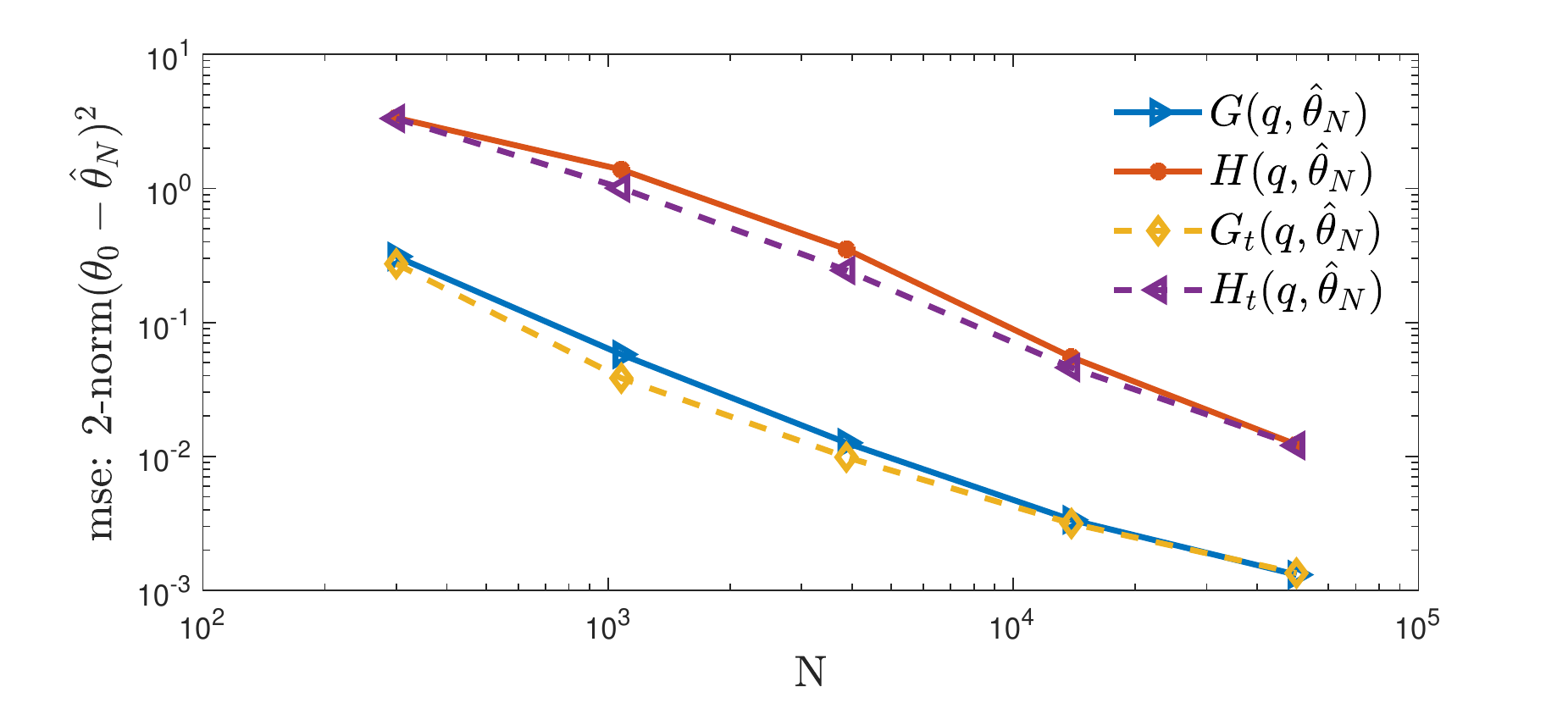}
  \caption{MSE between $\hat{\theta}_N$ and $\theta_0$ as function of sample size, averaged over the Monte Carlo runs, obtained with Algorithm 1 with $R^0=\begin{bmatrix}
      0& R_b^{0^\top}
  \end{bmatrix}^\top$ and $R_b=I\in \mathbb{R}^{2\times 2}$, where subscript $\{t\}$ indicates the use of the true (unknown) white noise as a predictor input instead of the reconstructed innovation.}
    \label{fig:reduced_rank_sim_MSE_r5_r6}
\end{figure}
%
%
Next we present the results of the estimation of the parametric model.
Because Algorithm 1 is consistent we have a negligible bias and the mean squared error (MSE) represents the variance. For the simulations we use the correct estimated disturbance topology from the previous step. Additionally, for Step 3.2 of Algorithm 1, we compute the $\hat{\theta}_{j_N}^{[0]}$ in \eqref{eq:WNSF_theta0} using the covariance matrix of the nonparametric model as weighting.
Figures \ref{fig:reduced_rank_sim_MSE} and \ref{fig:reduced_rank_sim_MSE_r5_r6} present the
sample MSE that is computed according to MSE$(N) = \frac{1}{M} \sum^{M}_{c=1} \begin{Vmatrix}
    \hat{\theta}_{N,c}-\theta_0
\end{Vmatrix}^2$, where $c$ indicates the Monte Calro run and $\hat{\theta}_{N,c}$ the final estimate \eqref{eq:WNSF_theta_end}. In Figure \ref{fig:reduced_rank_sim_MSE} we use $R^0=I$ in the data generating network, and in Figure \ref{fig:reduced_rank_sim_MSE_r5_r6} we use $R^0=\begin{bmatrix}
      0& R_b^{0^\top}
  \end{bmatrix}^\top$ with $R_b=I\in \mathbb{R}^{2\times 2}$ according to Section \ref{subsec:example_rr}. The solid lines represent Algorithm 1 where the estimates are obtained using the reconstructed innovation as input. The dotted lines represent Algorithm 1 where we use the realization of the actual noise $e(t)$ as input, indicated by subscript $\{t \}$. The results for the whole network are shown, while using $L$ linear regressions. %
  Both simulations shown in Figures \ref{fig:reduced_rank_sim_MSE} and \ref{fig:reduced_rank_sim_MSE_r5_r6}, typically perform $k=6$ iterations for data length $N=300$ in \eqref{eq:WNSF_theta_end}. As the data length $N$ increases the number of iterations performed decreases, where for $N=50000$ the simulations typically perform $k=2$ iterations. The MSE$(N)$ improvement after the iterations is shown in Table \ref{tb:update_k_mse}. From Table \ref{tb:update_k_mse} we can derive that we benefit most from iterating $k$ in the final step of Algorithm 1 if we do not have full excitation on the network with $R^0=I$.
\setlength{\tabcolsep}{3pt}
  \begin{table}[h]
\begin{center}
\caption{MSE improvement:\\ 
%
\blue{$ \frac{1}{M} \sum^{M}_{c=1} \begin{Vmatrix}
    \hat{\theta}_{N,c}- \theta_0
\end{Vmatrix}^2 -
\frac{1}{M} \sum^{M}_{c=1} \begin{Vmatrix}
    \hat{\theta}^{(1)}_{N,c}-\theta_0
\end{Vmatrix}^2$}
 over k iterations}\label{tb:update_k_mse}
\resizebox{\linewidth}{!}{
\begin{tabular}{c|ccccc}
 $N$ & 300  &1078 &3873 &13916 & 50000  \\
          \hline
%
$R^0\!=\! I$ &  $1.6\! \cdot10^{-3}$\! &$5.1\! \cdot10^{-5}$ \! & $-1.2\! \cdot10^{-6}$\! &$-1.9\! \cdot10^{-7}$\! &$3.7\! \cdot10^{-8}$\! \\
$R_b^0\!=\! I$ & 0.43 &0.26 & 0.15 & 0.07 & 0.01
\end{tabular}}
\end{center}
\vspace{-2pt}
\end{table}
%
%

In Figures \ref{fig:reduced_rank_sim_MSE} and \ref{fig:reduced_rank_sim_MSE_r5_r6} we see convergence between the solid and dotted lines as the data length $N$ increases. This indicates that as data length $N$ increases the reconstructed innovation converges to the actual noise.
Furthermore all MSE results continue to converge towards zero which is in line with the consistency proof.

The results of this simulation study support the consistency proof and we consistently estimate the BJ model structure, while \blue{employing a row-wise optimization.}
\section{Conclusions}\label{sec:conclusion}
In this paper we present a multi-step least squares method for network identification, that can handle reduced rank noise with low computational burden. We follow a step wise procedure where we first extend the SLR identification method to detect the disturbance topology, and thereafter extend the WNSF method to consistently identify networks of general model structure, including a BJ model structure. For a BJ network, usually a non-convex MIMO identification method is needed. In this paper, we show that we identify the BJ network using analytical solutions.
Simulation results indicate that we can identify the disturbance topology of the given network with low error if the data length $N$ is sufficiently large. We show that the presented method is consistent, and provide path based data informativity conditions, that guides where to allocate external excitation signals for the experimental design.
Considering large networks subject to correlated and/or reduced rank noise, the presented method is promising due to its scalability and low variance results.
\appendix
\section{Proof of Proposition \ref{prop:1}}
Consider the prediction error for the predictor $\hat w(t|t-1,\zeta)$ from \eqref{eq:pred_step1_rr_0}:
\begin{equation}
\begin{aligned}
    {\varepsilon} (t,\zeta)&= w(t)-\hat{w}(t|t-1,\zeta) = \breve{A}(\zeta) w(t) - \breve{B}(\zeta)r(t),\\
    &= \breve{A}(\zeta) w(t) - {B}(\zeta)r_a(t) - \begin{bmatrix}
    0 & R_b^\top    \end{bmatrix}^\top r_b (t).
\end{aligned}
\end{equation}
With the data generating system \eqref{eq:dyn_network1} given as
\begin{equation}
\begin{aligned}
    w(t)& = (\breve{A}^0)^{-1}\breve{B}^0  r(t) + (\breve{A}^0)^{-1} \breve{e}(t),\\
    & \mbox{with } \breve{A}^0=  (\breve H^0)^{-1}(I-G^0) , \ \ \breve{B}^0=  (\breve H^0)^{-1} R^0
\end{aligned}
\label{eq:AppA_w1}
\end{equation}
we can rewrite the prediction error as
\begin{equation}
   {\varepsilon} (t,\zeta)  = \big( \breve{A}^0-\Delta \breve{A}(\zeta) \big)  w - \big( \breve{B}^0-\Delta \breve{B}(\zeta) \big)  r
\end{equation}
with $\Delta \breve{A}(\zeta) = \breve{A}^0 - \breve{A} (\zeta)$ and $\Delta \breve{B}(\zeta) = \breve{B}^0 - \breve{B} (\zeta)$. Then with \eqref{eq:AppA_w1} it follows that
\begin{eqnarray}
{\varepsilon} (t,\zeta) & = & \Delta \breve{B}(\zeta) r  - \Delta \breve{A}(\zeta)  w + \breve e,
\end{eqnarray}
and since the second block column of $\breve{B}(\zeta)$ is fixed and known, it follows that $\Delta \breve{B}(\zeta) r = \Delta {B}(\zeta) r_a$.
We now proceed by evaluating the $j$-th component
\begin{equation}
\begin{aligned}
   & {\varepsilon}_{j} (t,\zeta) =
    &\Delta B_j(\zeta) r_a  - \Delta \breve{A}_j(\zeta) w
    + \breve{e}_j,
\end{aligned}
\label{eq:AppA_residual}
\end{equation}
where $\Delta\breve{A}_j(\zeta)$ and $\Delta{B}_j(\zeta)$ are the rows of matrices $\Delta\breve{A}(\zeta)$ and $\Delta{B}(\zeta)$ belonging to node $j$.\\
The consistency proof consists of two steps:
\begin{enumerate}
    \item Show that the objective function is bounded from below by the noise variance $\bar{V}_j (\zeta):= \bar{\mathbb{E}}\varepsilon_j^2(t,\zeta) \geq \sigma^2_{\breve{e}_j} $, where the minimum is achieved for $\Delta \breve{A}_j(\zeta) = 0$ and $\Delta \breve{B}_j(\zeta) = 0$.
    \item Show that the global minimum is unique.
\end{enumerate}
\subsection{Consistency proof step (1)}
With \eqref{eq:AppA_w1} substituted into \eqref{eq:AppA_residual}, the expression for $\varepsilon_j(t,\zeta)$ becomes
\begin{equation}
  \Delta B_j(\zeta)r_a   - \Delta \breve{A}_j(\zeta)  \Big( (\breve{A}^0)^{-1}\breve{B}^0 r  + (\breve{A}^0)^{-1} \breve{e} \Big) + \breve e_j
    \label{eq:Apendix_A_resid}
\end{equation}
from which, due to the fact that $\Delta \breve{A}_j(\zeta)$ is strictly proper and $r$ and $e$ are uncorrelated, it follows that $\breve e_j$ is uncorrelated with the remaining terms in the expression. As a result,
%
%
the objective function is given by
\begin{equation}
    \bar{V}_j(\zeta) = \bar{\mathbb{E}} \Big[ \Big(
      \Delta
      {B}_j(\zeta) r_a  - \Delta \breve{A}_j(\zeta)  w \Big)^2 \Big] +\sigma^2_{\breve{e}_j},
      \label{eq:V_proof1}
\end{equation}
from which we can infer that $\bar{V}_j(\zeta) \geq \sigma^2_{\breve{e}_j}$ with equality for $\Delta \breve A_j(\zeta) = 0$ and $\Delta B_j(\zeta) = 0$.
\subsection{Consistency proof step (2)}
For the second step we show that the minimum is unique, by showing that $\bar{V}_j (\zeta)= \sigma^2_{\breve{e}_j}$ implies $\Delta \breve A_j(\zeta) = 0$ and $\Delta B_j(\zeta) = 0$.
With \eqref{eq:V_proof1} and by applying Parseval's theorem, $\bar{V}_j (\zeta)= \sigma^2_{\breve{e}_j}$ implies
\begin{equation}
    \frac{1}{2\pi} \int ^\pi_{-\pi} \Delta x^\top (e^{j\omega},\zeta)^\top  \Phi_\kappa(\omega) 
    \Delta x(e^{-j\omega} , \zeta) \text{d}\omega = 0,
    \label{eq:A_parseval_w}
\end{equation}
with $\Delta x^\top = \begin{bmatrix}
      \Delta {B}_j(\zeta) &- \Delta \breve{A}_j(\zeta)
      \end{bmatrix}$ and $\kappa= \begin{bmatrix} r_a^\top & w^\top\end{bmatrix}^\top$.\\
%
%
%
%
%
By Condition (2) the spectral density $\Phi_\kappa(\omega)$ is positive definite.
Therefore equation \eqref{eq:A_parseval_w} holds only for $\Delta x^\top =0$ which is satisfied by Condition (3). The global minimum of $\bar{V}_j(\zeta)$ is thus unique for $\breve{A}_j(\zeta)=\breve{A}^0_j$ and
$\begin{bmatrix} {B}_j(\zeta) & \begin{matrix}  \bar R_j \end{matrix}\end{bmatrix}=\breve{B}^0_j$, with $\bar R_j=0$ for $j=1,\dots,p$ and $\bar R_j$ is a row of $ R_b$ for $j=p+1,\dots,L$. \hfill $\qed$

%
%
\section{Proof of Proposition \ref{prop:2}}
%
%
%
%
For ease of notation we start with the MIMO notation of the one-step-ahead predictor \eqref{eq:pred_step2_MISO}
\begin{equation}
      \hat{w}(t|t-1,\eta) =  G(\eta) w + Rr+ 
      \bar{H}(\eta) {{\varepsilon}_a}(\hat{\zeta}^n_N) ,
\end{equation}
From Proposition 1 we know $\hat{\zeta}^n_N$ is consistent, therefore
\begin{equation}
\begin{aligned}
     {\varepsilon} (\hat{\zeta}^n_N) \to \breve{e}
     \quad \text{w.p. 1 as} \, N\to \infty \, \forall t,
\end{aligned}
\end{equation}
and we can rewrite the one-step-ahead predictor as
\begin{equation}
      \hat{w}(t|t-1,\eta) =  G(\eta) w + Rr+ 
      \bar{H}(\eta) {e}
\end{equation}
Considering the data generating system in \eqref{eq:dyn_network1} the residual becomes
\begin{equation}
\begin{aligned}
    {\varepsilon} (t,\eta)
    &=  w(t)-\hat{w}(t|t-1,\eta)\\
    &=
     \Delta G(\eta) w + H^0 e - \bar{H}(\eta) e\\
    &=
    \Delta G(\eta) w + \Delta \bar{H}(\eta) e + \begin{bmatrix} I \\ \Gamma^0 \end{bmatrix} e,
 \end{aligned}
\end{equation}
where $\Delta G(\eta)= G^0 - G(\eta)$, and
$
    \Delta \bar{H}(\eta)= \begin{bmatrix} \Delta \bar{H}_a(\eta) \\ \Delta \bar{H}_b(\eta) \end{bmatrix},
$
with
$\Delta \bar{H}_a(\eta)=\bar{H}_a^0 - \bar{H}_a(\eta)$, with $\bar{H}_a=H_a-I$ and $\Delta \bar{H}_b(\eta)= \bar{H}_b^0 - \bar{H}_b(\eta)$, with $\bar{H}_b= H_b - \Gamma$.\\
The residual per node is written as
\begin{equation}
\begin{aligned}
   {\varepsilon}_j(t,\eta) 
   &=
   \sum_{\substack{l\in \mathcal{N}_j}} 
   \Delta G_{jl}(\eta) w_l + 
   \sum_{s\in \mathcal{V}_j} \Delta \bar{H}_{js}(\eta)e_s +\breve{e}_j,
\end{aligned}
\label{eq:xxrs}
\end{equation}
where $\Delta G_{jl}(\eta) = G^0_{jl}-G_{jl}(\eta)$ is an element of matrix $\Delta {G}(\eta)$, 
and $\Delta \bar{H}_{js}(\eta)$ is an element of matrix $\Delta \bar{H}(\eta)$.\\
The consistency proof consists of two steps
\begin{enumerate}
    \item 
    Show that the objective function is bounded from below by the noise variance\footnote{\blue{$ \overline{\mathbb{E}}$ refers to the generalized expectation operator $\lim_{N\to \infty} \frac{1}{N} \sum^{N}_{t=1} \mathbb{E}$.}} $\bar{V}_j (\theta):= \bar{\mathbb{E}}\varepsilon_j^2(t,\theta) \geq \sigma_{\breve{e}_j}^2 $, where the minimum is achieved for $\Delta G_{jl}=0$ 
    and $\Delta \bar{H}_{js}=0$.
    \item Show that the global minimum is unique. 
\end{enumerate}
\emph{Step 1}
By using the property that all $\Delta G$- and $\Delta \bar H$-terms are strictly proper, it follows from \eqref{eq:xxrs} that
\begin{equation}
\begin{aligned}
    \bar{V}_j(\eta) &= \bar{\mathbb{E}} \Big[ \Big(
    \sum_{\substack{l\in \mathcal{N}_j}} 
   \Delta G_{jl}(\eta) w_l  
   +
    \sum_{s\in \mathcal{V}_j} \Delta \bar{H}_{js}(\eta)e_s
   \Big)^2 \Big]+ \sigma_{\breve{e}_j}^2
\end{aligned}
\label{eq:V_bar_proof2}
\end{equation}
and $\bar{V}_j(\eta) \geq \sigma^2_{\breve{e}_j}$ with equality for $\Delta G_{jl} = 0$ and $\Delta \bar H_{js} = 0$ for all $l\in \mathcal{N}_j$ and $s\in \mathcal{V}_j$.

\emph{Step 2}
Showing that the minimum is unique is done by showing that $\bar{V}_j (\eta)= \sigma^2_{\breve{e}_j}$ implies $\Delta G_{jl} = 0$ and $\Delta \bar H_{js} = 0$ for all $l\in \mathcal{N}_j$ and $s\in \mathcal{V}_j$. With \eqref{eq:V_bar_proof2} and by applying Parseval's theorem, $\bar{V}_j (\zeta)= \sigma^2_{\breve{e}_j}$ implies
\begin{equation}
       \frac{1}{2\pi} \int ^\pi_{-\pi} \Delta x^\top (e^{j\omega},\eta)^\top  \Phi_{\bar{\kappa}}
       (\omega)  \Delta x(e^{-j\omega} , \eta) \text{d}\omega = 0 ,
       \label{eq:proof_parseval_B_w}
\end{equation}
with $\Delta x^\top\! =\!  \begin{bmatrix} \Delta G_{jl\in \mathcal{N}_j} &%
    \Delta  \bar{H}_{js\in \mathcal{V}_j}
    \end{bmatrix}$ and $\bar{\kappa}= \begin{bmatrix} w_{\{ \mathcal{N}_j\}}^\top & e_{\{ \mathcal{V}_j\}}^\top \end{bmatrix}^\top\! .$\\
By Condition (1) the spectral density $\Phi_{\bar{\kappa}}$ is positive definite. Therefore equation \eqref{eq:proof_parseval_B_w} holds only for $\Delta x^\top = 0$.
The Parseval's theorem shows the the global minimum of $\bar{V}_j(\eta)$ is unique for $G_{jl}(\eta)=G^0_{jl}$ and $\bar{H}_{js}(\eta)= \breve H^0_{js}-I_{js}$ by Condition (2). \hfill $\qed$
\section{Proof of Proposition \ref{prop:3}}
The vector signal $\kappa$ is written as
\begin{equation}
\kappa = \begin{bmatrix} r_a \\ w \end{bmatrix} = \underbrace{\begin{bmatrix} I & 0 & 0 \\ J_{wa} & J_{wb} & J_{we} \end{bmatrix}}_{J} \begin{bmatrix} r_a \\  r_b \\ e \end{bmatrix}
\end{equation}
with $J_{wa}, J_{wb}, J_{we}$ appropriate transfer function matrices. Since $\rho= \begin{bmatrix} r_a^\top & r_b ^\top&  e^\top        \end{bmatrix} ^\top$ is persistently exciting, i.e. $\Phi_{\rho}(\omega) \geq 0$ for all $\omega$, it follows from Lemma 1 in \cite{Hof2020path} that $\kappa$ is persistently exciting if and only if matrix $J$ has full row rank. Since full row rank of $J$ is equivalent to a full row rank of $[J_{wb}\ J_{we}]$, the result of Proposition 1 in \cite{Hof2020path} then shows the equivalence with the condition that there are $L$ vertex disjoint paths from the inputs of $[J_{wb}\ J_{we}]$, i.e. $r_b$ and $e$,  to its outputs, i.e. $w$. \hfill $\qed$
\section{Proof of Proposition \ref{prop:4}}
Similar to the line of reasoning in the proof of Proposition \ref{prop:3}, the vector signal $\bar\kappa$ is written as
\begin{equation}
\bar\kappa = \begin{bmatrix} w_{\{ \mathcal{N}_j\}} \\ e_{\{\mathcal{V}_j\}} \end{bmatrix} = \underbrace{\begin{bmatrix} J_{wr} & J_{wx} & J_{wv} \\ 0 & 0 & I \end{bmatrix}}_{\bar J} \begin{bmatrix} r\\ e_{\{\mathcal{X}_j\}} \\  e_{\{\mathcal{V}_j\}} \end{bmatrix}
\end{equation}
with $J_{wr}, J_{wx}, J_{wv}$ appropriate transfer function matrices. Since $\bar \rho= \begin{bmatrix} r^\top &    e_{\{\mathcal{X}_j\}} ^\top&  e_{\{\mathcal{V}_j\}}^\top \end{bmatrix} ^\top$ is persistently exciting, i.e. $\Phi_{\bar\rho}(\omega) \geq 0$ for all $\omega$, it follows from Lemma 1 in \cite{Hof2020path} that $\bar\kappa$ is persistently exciting if and only if matrix $\bar J$ has full row rank. Since full row rank of $\bar J$ is equivalent to a full row rank of $[J_{wr}\ J_{wx}]$, the result of Proposition 1 in \cite{Hof2020path} then shows the equivalence with the condition that there are $Cardinal\{\mathcal{N}_j\}$ vertex disjoint paths from the inputs of $[J_{wr}\ J_{wx}]$, i.e. $r$ and $e_{\{\mathcal{X}_j\}}$,  to its outputs, i.e. $w_{\{ \mathcal{N}_j\}}$. \hfill $\qed$
%
%
\section{System used in simulations}\label{App:used_models}
In the simulation results in Section \ref{sec:Num_ex} we use the data generating network \blue{of which the graph is} represented in Figure \ref{fig:6node_fig}. The data generating transfer functions $G$ and $H$ are given by
%
\begin{equation}
   G = \begin{bsmallmatrix}
    0 & 0 & 0 & G_{14} & 0 & 0\\
     0 & 0 & 0 & 0 &  G_{25}& 0\\
      G_{31} & 0 & 0 & 0 & G_{35} & 0\\
       0 & G_{42} & 0 & 0 & 0 & 0\\
        G_{51} & 0 & 0 & 0 & 0 & G_{56}\\
         0 & 0 & G_{63} & 0 & 0 & 0\\
    \end{bsmallmatrix},
\end{equation}
with the elements of $G_{jl}$
\begin{equation}
\begin{aligned}
&G_{14} = \tfrac{ 0.38q^{-1} + 0.24 q^{-2} }{ 1- 1.35 q^{-1} +0.54 q^{-2}}
,
&G_{25}= \tfrac{ 0.20 q^{-1}  }{ 1- 1.30 q^{-1} +0.60 q^{-2}}
,\\
&G_{31}= \tfrac{ 0.39 q^{-1}  }{ 1- 0.80 q^{-1} +0.20 q^{-2}}
,
&G_{35}= \tfrac{ 0.16 q^{-1}  }{ 1- 1.23 q^{-1} +0.51 q^{-2}}
,\\
&G_{42}= \tfrac{ -0.30 q^{-1}  }{ 1- 0.60 q^{-1} +0.20 q^{-2}}
,
&G_{51}= \tfrac{ -0.60 q^{-1}  }{ 1+0.45 q^{-1} +0.12 q^{-2}}
,\\
&G_{56}= \tfrac{ -0.22 q^{-1}  }{ 1- 1.22 q^{-1} +0.46 q^{-2}}
,
&G_{63}= \tfrac{ -0.11 q^{-1}  }{ 1- 1.49 q^{-1} +0.62 q^{-2}},
\end{aligned}
\end{equation}
and
\begin{equation}
\begin{aligned}
H= \begin{bsmallmatrix}
    H_{11} & 0 & 0 & H_{14} \\
     0 & H_{22} & 0 & 0 &  \\
      0 & H_{32} & H_{33} & 0 \\
       0 & H_{42} & 0 & H_{44} \\
        0 & H_{52} & H_{53} & 0 \\
         0 & H_{62} & 0 & H_{64} \\
\end{bsmallmatrix},
\end{aligned}
\end{equation}
with noise rank $p=4$ and elements
\begin{equation}
\begin{aligned}
&H_{11} = \tfrac{1+ 0.52 q^{-1}  }{ 1+ 0.41 q^{-1} }
,
&H_{14}= \tfrac{ 0.41 q^{-1}  }{ 1 -0.56 q^{-1} }
,\\
&H_{22}= \tfrac{1+ 0.44 q^{-1}  }{ 1+ 0.35 q^{-1} }
,
&H_{32}= \tfrac{ -0.56 q^{-1}  }{ 1-0.40  q^{-1} }
,\\
&H_{33}= \tfrac{1-0.20  q^{-1}  }{ 1+0.43  q^{-1} }
,
&H_{42}= \tfrac{ 0.26 q^{-1}  }{ 1-0.62  q^{-1} }
,\\
&H_{44}= \tfrac{1+ 0.52 q^{-1}  }{ 1+0.45  q^{-1} }
,
&H_{52}= \tfrac{0.49  q^{-1}  }{ 1-0.49 q^{-1} }
,\\
&H_{53}= \tfrac{1+ 0.66  q^{-1}  }{ 1+ 0.51  q^{-1} }
,
&H_{62}= \tfrac{1+ 0.24 q^{-1}  }{ 1+ 0.53 q^{-1} }
,\\
&H_{64}= \tfrac{-0.56  q^{-1}  }{ 1-0.56 q^{-1} +0.21 q^{-2} },
\end{aligned}
\end{equation}
where $\Gamma^0= \begin{bmatrix}
    0&0&1&0\\
    0&1&0&0
\end{bmatrix}$.
%
%
%
%
%
%
\bibliographystyle{plain}        
\bibliography{references}           


%
\end{document}